\documentclass[conference]{IEEEtran}
\usepackage{amssymb}

\usepackage[english]{babel}
\usepackage[utf8]{inputenc}
\usepackage{amsmath}
\usepackage{graphicx}
\usepackage{hyperref}
\usepackage{algpseudocode}
\usepackage[colorinlistoftodos]{todonotes}
\usepackage{enumitem}
\usepackage{array}
\usepackage{changepage}
\usepackage{longtable}
\usepackage{pdflscape}
\usepackage{booktabs}
\usepackage{float}

\usepackage{epstopdf}
\usepackage[ruled,vlined]{algorithm2e}
\usepackage{tabularx}

\newpage
\title{Cooperative Flexibility Exchange: Fair and Comfort-Aware Decentralized Resource Allocation}

\author{%
	Rabiya Khalid$^{*}$ and Evangelos Pournaras\\
	School of Computer Science, University of Leeds, Leeds, United Kingdom\\
	Email: \{r.khalid, e.pournaras\}@leeds.ac.uk
	\thanks{$^{*}$Corresponding author: r.khalid@leeds.ac.uk}
}
\begin{document}
	\maketitle
	\begingroup
	\renewcommand\thefootnote{*}
	\endgroup
	\begin{abstract}
		
		The growing electricity demand and use of smart appliances are placing pressures on power grids, making efficient energy management more important than ever. 
		The existing energy management systems often prioritize system efficiency (balanced energy demand and supply) at the expense of consumer comfort. This paper addresses this gap by proposing a novel decentralized multi-agent coordination-based demand-side management system. The proposed system enables individual agents to coordinate for demand-side energy optimization while improving the consumer comfort and maintaining the system efficiency. A key innovation of this work is the introduction of a slot exchange mechanism, where agents first receive optimized appliance-level energy consumption schedules and then coordinate with each other to adjust these schedules through slot exchanges to improve their comfort even when agents show non-altruistic behaviour. It also scales well with large populations and promotes fairness by balancing satisfaction levels across consumers. For performance evaluation, a real-world dataset is used, and the results demonstrate that the proposed slot exchange mechanism increases consumer comfort and fairness without raising system inefficiency cost, making it a practical and scalable solution for future smart grids.
	\end{abstract}
	\noindent\textbf{Keywords:} Smart Grids; Demand-Side Management; Energy Management System; Decentralized Optimization; consumer Comfort; Fairness; Multi-Agent Systems; Cooperative Flexibility Exchange
	
	\section{Introduction}
	\begin{figure}
		\centering
		\includegraphics[width=0.9\linewidth]{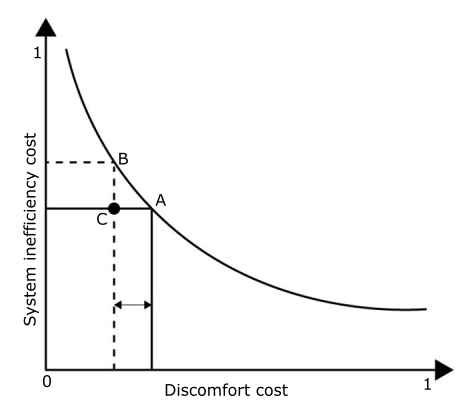}
		\caption{Trade-off between consumer discomfort cost (x-axis) and system inefficiency cost (y-axis) in residential demand-side management. Point A corresponds to a solution with high discomfort and inefficiency costs. Moving towards point B reduces consumer discomfort at the expense of increased system inefficiency, illustrating the conflicting nature of the two objectives. Point C highlights a desirable operating region where consumer discomfort is reduced without increasing system inefficiency, which motivates the coordination mechanism proposed in this work.}
		
		\label{fig:picture1}
	\end{figure}
	
	In smart grids, demand-side management plays a crucial role in balancing electricity demand with supply, particularly given the increasing penetration of renewable energy sources and the growing variability of residential electricity consumption. Renewable energy sources are projected to account for nearly 45\% of global electricity generation by 2030 \cite{IEA2030}, significantly increasing supply-side uncertainty and reinforcing the need for flexible and coordinated demand response (DR) at the residential level.
	
	Residential demand-side management has therefore become a key component of modern smart grids, typically implemented through home energy management systems (HEMSs) that schedule the operation of flexible household appliances, such as washing machines, dishwashers, and electric vehicle charging, over discrete time slots. While extensive research has focused on optimizing these schedules to reduce peak demand and system-level costs, the explicit consideration of consumer comfort has received comparatively limited attention \cite{intro1}. In many existing approaches, comfort is treated as a secondary constraint or as a trade-off against grid performance, which can negatively affect consumer acceptance and long-term participation in demand-side management programs.
	
	In this work, consumer comfort is defined in a schedule-based manner, reflecting deviations from consumers’ preferred appliance operation times. Such deviations, caused by advancing or delaying appliance execution, are modeled as a quantifiable discomfort cost. Enhancing comfort therefore corresponds to minimizing these schedule deviations while preserving system-level objectives, such as maintaining a balanced electricity demand and supply. This formulation allows comfort to be explicitly incorporated into the optimization process rather than being implicitly sacrificed for efficiency gains.
	
	To address the decentralized and heterogeneous nature of residential demand-side management, this paper adopts a multi-agent system (MAS) framework. In this setting, each household is represented by an autonomous agent associated with a local HEMS controller. Agents independently determine appliance schedules based on local comfort preferences and operational flexibility, while interacting with other agents to collectively satisfy grid-level constraints. Through coordination, agents can adjust their schedules in a distributed manner, enabling system-wide demand balancing without requiring centralized control or full information sharing.
	
	Multi-agent coordination has proven effective in various large-scale, decentralized domains, including traffic management \cite{trafic,trapp,vehicle} and supply chain systems \cite{supply,supply2}, where local decision-making combined with coordination leads to improved system level efficiency. In the context of demand-side management, existing multi-agent approaches primarily focus on reducing energy costs or peak demand using decentralized coordination \cite{MADC}, reinforcement learning \cite{MARL,d5}, or load shifting techniques \cite{MALS}. However, these methods are largely driven by economic and environmental objectives, with limited emphasis on consumer comfort and its distribution across participants.
	
	A fundamental challenge in residential demand-side management lies in the inherent trade-off between consumer comfort and system-level efficiency. Reducing system inefficiency typically requires shifting or rescheduling flexible loads to align demand with available supply, particularly during periods of high renewable generation or peak demand. However, such adjustments often deviate from consumers’ preferred appliance operation schedules, resulting in increased discomfort. Conversely, strictly preserving consumer comfort by adhering to preferred schedules can limit the system’s flexibility and lead to higher aggregate inefficiency, such as increased peak loads or imbalanced demand and supply. This trade-off is intrinsic to demand-side management and motivates the need for mechanisms that can reduce consumer discomfort without degrading system-level performance.
	
	Beyond discomfort cost, fairness is another critical yet underexplored aspect of residential demand-side management. Fairness can be understood as the equitable distribution of discomfort among agents. Without explicit mechanisms to account for fairness, some households may repeatedly experience higher discomfort due to limited flexibility or unfavorable scheduling outcomes \cite{unfair,flexibility}, undermining trust and willingness to cooperate. Additionally, residential agents may exhibit different degrees of altruism or selfishness, prioritizing local comfort over collective goals to varying extents \cite{altruism}. These behavioral differences can significantly affect system performance and stability.
	
	In a residential setting, each household naturally prioritizes its own comfort and convenience, which may conflict with collective grid-level objectives. Selfish agents tend to favor schedules that minimize their individual discomfort, even if such decisions negatively impact overall system balance. In contrast, altruistic agents are willing to accept higher local discomfort to support global objectives, such as reducing peak demand or accommodating renewable generation. In practice, residential agents exhibit varying degrees of altruism, shaped by individual preferences, flexibility, and willingness to cooperate. Understanding how different levels of altruistic and selfish behavior affect system performance, consumer comfort, and fairness is therefore essential for designing robust and realistic demand-side management coordination mechanisms.
	
	To address these challenges, this paper proposes a novel multi-agent coordination mechanism that explicitly enhances consumer comfort and fairness while maintaining the balance between electricity demand and supply. The proposed approach operates in two stages. First, agents coordinate to obtain a globally feasible schedule that satisfies grid-level constraints while accounting for individual comfort preferences through multi-objective optimization. Second, agents engage in localized coordination to further improve comfort by adjusting schedules, without compromising the achieved system-wide balance. Through this design, the proposed framework moves beyond traditional demand-side management approaches by embedding comfort and fairness as core objectives rather than secondary considerations.

	\begin{itemize}
		\item The design of a novel comfort-aware, decentralized agent-based demand-side management mechanism, in which agents collaboratively adjust their energy consumption schedules through a coordination-based slot exchange process to improve consumer comfort while preserving grid stability.
		
		\item A comprehensive study of the impact of agent altruism and proposed slot exchange protocol on comfort and fairness in the system.
		\item The demonstration of the approach in realistic smart grid scenarios, validating its effectiveness and practical applicability.
		\item The release of open-source code and dataset to facilitate reproducibility and support further research in decentralized, comfort-aware demand side management.
	\end{itemize}
	This work advances the state of the art in demand side management research towards more consumer-centric, equitable, and resilient solutions, paving the way for next-generation smart grids that align technological efficiency with human comfort and social acceptance. Beyond academic contributions, the findings are directly relevant for energy communities, which can adopt the proposed mechanism to foster cooperative and fair energy sharing; for system operators, who can leverage the approach to maintain grid stability while improving consumer participation; and for policy makers, who can use the insights to design incentive structures and regulatory frameworks that balance efficiency with fairness and consumer well-being.

	The rest of the paper is organized as follows. Section \ref{RW} reviews the related work. Section \ref{PF} formulates the problem, outlining the specific challenges and objectives that guide the study. In Section \ref{PS}, we propose a system model and solution to the identified problem. Section \ref{SR} presents the experimental results. Sections \ref{flim} outline the limitation and futurework. Finally, Section \ref{con} concludes the paper.
	\begin{table*}[htbp]
		\centering
		\caption{Comparative summary of related studies }
		\small
		\begin{tabular}{@{} c  p{10cm}  c c c c@{}}
			\toprule
			\textbf{Ref} & \textbf{Focus Area} & \textbf{MAS} & \textbf{Comfort} & \textbf{Fairness} & \textbf{Altruism} \\
			\midrule
			\cite{MAS}   & Smart grid energy loss minimization & $\checkmark$ & $\times$
			& $\times$
			& $\times$
			\\
			\cite{2}     & Multi-agent coordination& $\checkmark$ & Partial & $\times$
			& $\times$
			\\
			\cite{MADC}  & Microgrids & $\checkmark$ & $\times$
			& $\times$
			& $\times$
			\\
			\cite{MARL}  & Home energy management & $\checkmark$ & Partial & $\times$
			& $\times$
			\\
			\cite{MALS}  & Demand side management across sectors (Antlion optimizer) & $\checkmark$ & $\times$
			& $\times$
			& $\times$
			\\
			\cite{rw1}   & Energy demand (multi-agent deep reinforcement learning) & $\checkmark$ & Partial & $\times$
			& $\times$
			\\
			\cite{rw3}   & DR (game-theoretic) & $\times$
			& Partial & $\times$
			& $\times$
			\\
			\cite{rw4}   & Demand side management energy sharing (prosumers) & $\checkmark$ & Partial & $\times$
			& $\times$
			\\
			\cite{rw5}   & Peak shaving (transfer learning) & $\checkmark$ & $\times$
			& $\times$
			& $\times$
			\\
			\cite{rw6}   & Economic dispatch (MAS + PSO) & $\checkmark$ & $\times$
			& $\times$
			& $\times$
			\\
			\cite{rw7}   & Two-stage robust framework & $\times$
			& $\times$
			& $\times$
			& $\times$
			\\
			\cite{rw8}   & Hybrid microgrids DR (MAS) & $\checkmark$ & $\times$
			& $\times$
			& $\times$
			\\
			\cite{com1}  & Smart home (GA vs PSO) & $\times$
			& $\checkmark$     & $\times$
			& $\times$
			\\
			\cite{com2}  & Visual comfort (negotiation) & $\times$
			& $\checkmark$     & $\times$
			& $\times$
			\\
			\cite{com3}  & Thermal comfort (multi-objective) & $\times$
			& $\checkmark$     & $\times$
			& $\times$
			\\
			\cite{com4}  & IAQ \& comfort (Modified Bat) & $\times$
			& $\checkmark$     & $\times$
			& $\times$
			\\
			\cite{com5}  & Smart home AI (hybrid AI) & $\times$
			& $\checkmark$     & $\times$
			& $\times$
			\\
			\cite{com6}  & Hydrogen homes& $\times$
			& Partial & $\times$
			& $\times$
			\\
			\cite{farzam} & Appliance-level flexible scheduling & $\checkmark$ & Partial & $\checkmark$ & $\checkmark$ \\
			\cite{rr1} & EV fast-charging pricing with power--traffic coupling & $\checkmark$ & $\times$ & $\checkmark$ & Partial  \\
			\cite{rr2} & Two-stage EV charging with Stackelberg game & $\checkmark$ & $\times$ & $\times$ & $\times$ \\
			\cite{rr3} & Behavioral flexibility in energy communities & $\times$ & $\checkmark$ & Partial & $\checkmark$ \\
			\cite{rr4} & Aggregator-based flexibility coordination & $\checkmark$ & $\times$ & Partial & $\times$ \\
			\cite{rr5} & Review of BTM flexibility, peer-t-peer trading, machine learning approaches & $\times$ & Partial  & Partial  & $\times$ \\
			\cite{rr6} & Commercial peer-t-peer energy sharing with demand flexibility & $\checkmark$ & Partial  & Partial  & $\times$ \\
			\cite{rr7} & Industrial demand side management with peer-t-peer trading via multi-agent reinforcement learning & $\checkmark$& $\times$ & $\times$ & $\times$ \\
			\cite{rr8} & Peer-t-peer electricity usage-rights (quota) transfer & $\checkmark$& $\times$ & $\checkmark$ & Partial  \\
			\cite{rr9} & Review of robust reinforcement learning for peer-t-peer DR & $\times$ & $\times$ & $\times$ & $\times$ \\
			\cite{rr10} & Cloud resource scheduling with energy constraints & Partial  & $\times$ & $\times$ & $\times$ \\
			\cite{rr11} & Heuristic scheduling under continuous energy constraints & $\times$ & $\times$ & $\times$ & $\times$ \\
			\cite{rr12} & Three-layer game-theoretic microgrid scheduling & $\checkmark$& $\times$ & Partial  & $\times$ \\
			\cite{rr13} & Multi-timescale Stackelberg DR under uncertainty & $\checkmark$ & $\times$ & $\times$ & $\times$ \\
			\cite{rr14} & Privacy-preserving noncooperative energy management & $\checkmark$& $\times$ & $\times$ & $\times$ \\
			\cite{rr16} & Noncooperative microgrid planning under imperfect information & $\checkmark$ & $\times$ & $\times$ & $\times$ \\
			\cite{rr17} & DR cooperative minimizing demand charges & $\checkmark$ & $\times$ & $\checkmark$ & Partial  \\
			\cite{rr19} & Iterative price-based distributed DR & $\times$  & Partial  & $\times$ & $\times$ \\
			\midrule
			
			\textbf{Proposed} & Comfort-aware demand side management with slot exchange & $\checkmark$ & $\checkmark$ & $\checkmark$ & $\checkmark$ \\
			
			\bottomrule
		\end{tabular}
		\label{tab:comparative_related}
	\end{table*}
	\section{Related work}\label{RW}
	This section reviews existing work on demand-side management and multi-agent-based energy coordination. A comparative summary of the most relevant approaches is provided in Table \ref{tab:comparative_related} to highlight key differences and limitations. In the comfort column, partial refers to cases where comfort is addressed only indirectly or to a limited extent (for example, comfort may be treated as a constraint or tradeoff).
	
	\subsection{MAS-based energy management}
	The literature presents a wide range of agent-based and learning-driven approaches for energy management in smart grids and microgrids. Several studies employ multi-agent systems (MAS) to minimize energy losses, operational costs, and peak demand through decentralized coordination, storage optimization, and demand shifting strategies \cite{MAS,MALS}. Graph-based coordination methods and decentralized multi-agent energy management systems have also been proposed to enhance robustness, fault tolerance, and inter-microgrid power sharing under both normal and failure conditions \cite{2,MADC,rw2}.
	
	Learning-based approaches, including multi-agent reinforcement learning and deep reinforcement learning, have been applied to residential and microgrid energy management to balance energy demand and supply while considering electricity costs and user preferences \cite{MARL,rw1}. Game-theoretic and evolutionary optimization frameworks further model consumer dissatisfaction or payoff functions to guide DR and load shifting from peak to off-peak periods \cite{rw3}. In addition, several studies explore coordinated energy sharing among prosumers, dynamic pricing mechanisms, and DR strategies to improve grid sustainability, reduce emissions, and increase profitability \cite{rw4,rw5}.
	
	For interconnected and hybrid microgrids, MAS-based optimization frameworks have been introduced to improve distributed energy resource utilization and market participation \cite{rw8}. Incentive-driven and two-stage optimization approaches have also been explored for economic dispatch and demand side management across heterogeneous buildings and agent types, demonstrating improved efficiency and energy savings \cite{rw6,rw7}. Overall, these works highlight the effectiveness of decentralized, agent-based coordination for improving grid efficiency, cost reduction, and resilience, while largely focusing on economic and operational objectives rather than explicitly optimizing user comfort and fairness.

	\subsection{Optimizing consumer comfort}

	Recent advances in smart home technologies have increasingly focused on energy-efficient solutions that balance energy consumption with user comfort. Existing studies primarily model comfort through environmental factors such as thermal comfort, visual comfort, and indoor air quality. Thermal comfort relates to occupants’ satisfaction with temperature and humidity, visual comfort concerns lighting adequacy, and indoor air quality reflects the cleanliness and freshness of indoor air, directly affecting health and well-being.
	
	Emerging solutions also explore the integration of advanced optimization with sustainable energy technologies, such as hydrogen-based storage systems, to manage the trade-off between energy costs and comfort more effectively \cite{com6}. Overall, these studies demonstrate the effectiveness of comfort-aware optimization in smart homes, while largely focusing on environmental comfort variables rather than coordination-based comfort improvements through energy scheduling. Addressing multi-occupant environments, negotiation-based and soft-computing approaches have been proposed to reconcile diverse comfort preferences, particularly for visual comfort and lighting control \cite{com2}. Multi-objective optimization techniques have also been applied to thermal comfort, balancing energy costs with occupant satisfaction across varying climatic conditions \cite{com3}.
	
	Prior studies reveal a persistent tradeoff between energy efficiency and consumer comfort in smart homes \cite{com4}. Comfort, shaped by thermal conditions, lighting, and air quality, has been improved using optimization techniques such as genetic algorithm ()GA), particle swarm optimization (PSO), bat algorithm (BA), and hybrid artificial inteligence (AI) models \cite{com1}. However, cost reduction often requires appliance delays, which negatively affect user convenience. Despite continued advances in comfort-oriented optimization \cite{com5}, the tension between energy savings and comfort remains unresolved.
	
	Most demand side management strategies address comfort by regulating HVAC and lighting systems \cite{epsrc1}, yet cost savings are frequently achieved at the expense of user comfort \cite{epsrc2}. To compensate, monetary incentives such as bill credits and dynamic pricing are widely used \cite{comR1}. Auction-based coordination \cite{newr1}, optimization-driven incentives \cite{newr2}, and reward-based DR schemes \cite{newr3,newr4} show that financial rewards can increase participation, but engagement may decline when discomfort is frequent \cite{comR2}. Multi-objective optimization approaches attempt to balance comfort, cost, and peak-to-average load ratio, though fixed weightings limit their ability to capture evolving user preferences \cite{epsrc3}. 
	
	\subsection{Cooperative scheduling and peer coordination in demand side management}
	
	A substantial body of demand side management literature focuses on coordinating flexible demand to balance electricity supply and demand while reducing system costs or peak load. Early and widely adopted approaches rely on iterative price-based mechanisms, where a coordinator updates electricity prices and consumers independently reschedule their loads until convergence to a welfare-optimal or near-optimal equilibrium \cite{rr19}. These methods achieve system-level efficiency through implicit coordination but do not enable direct interaction among consumers, and improvements in user comfort are typically indirect, arising from relaxed constraints or price incentives rather than explicit coordination.
	
	Beyond price-based methods, cooperative and community-based demand side management frameworks have been proposed in which groups of consumers, prosumers, or microgrids jointly optimize energy consumption, often with centralized or hierarchical optimization followed by settlement or cost-sharing mechanisms \cite{rr17,rr12,rr1}. While cooperation can significantly reduce costs and demand charges, individual users typically have limited influence over the resulting schedules, and comfort is usually treated as a soft constraint rather than a primary objective.
	
	More recently, peer-to-peer coordination mechanisms have emerged that allow users to exchange energy, flexibility, or usage rights. Examples include P2P energy trading, quota or tariff-rights transfer, and matching-based allocation schemes (\cite{rr8,rr3,rr9,rr5}. These mechanisms move closer to decentralized coordination by introducing explicit peer interaction. However, most existing P2P approaches trade energy quantities or abstract rights, such as monthly quota or flexibility capacity, rather than discrete time-indexed service opportunities. Moreover, many of these methods primarily target economic efficiency or billing fairness, while user discomfort and behavioral heterogeneity are not explicitly optimized.
	
	It is important to distinguish the proposed slot exchange mechanism from existing cooperative scheduling or heuristic correction methods. Local correction heuristics in iterative demand side management typically adjust schedules after optimization to restore feasibility or reduce violations, without modeling bilateral acceptance, incentives, or systematic peer interaction. In contrast, slot exchange operates as an explicit coordination layer in which agents exchange discrete consumption slots to improve individual comfort while respecting global inefficiency constraints. The exchanged object is therefore directly tied to user experience and scheduling feasibility, rather than being an indirect or abstract representation of flexibility.
	
	\subsection{Game-theoretic models for decentralized energy management }
	
	Game theory has been extensively applied to demand side management to model strategic interactions among system operators, aggregators, and consumers. Stackelberg and bilevel game formulations are commonly used to represent hierarchical decision-making, where a leader sets prices or incentives and followers respond by optimizing their own objectives \cite{rr2,rr13}. These models provide strong theoretical guarantees and are well suited for regulatory or market-driven settings, but they primarily influence behavior through pricing signals and typically do not support peer-level schedule coordination.
	
	Cooperative game-theoretic approaches address fairness and stability by forming coalitions and allocating joint benefits using mechanisms such as Shapley value or negotiated settlements \cite{rr1,rr17}. While effective in ensuring that participants benefit from cooperation, these approaches usually assume centralized computation of the optimal schedule and focus on payoff allocation rather than on the structure of individual schedules or user comfort.
	
	Noncooperative game formulations model privacy-preserving or fully decentralized decision-making, often proving the existence and uniqueness of Nash equilibria under dynamic pricing \cite{rr14,rr16}. These models capture strategic behavior and information constraints but again focus on equilibrium consumption quantities rather than on explicit coordination of time-indexed activities.
	
	
	The present work connects to the game-theoretic literature by interpreting decentralized coordination as an iterative multi-agent optimization process with implicit incentives. The proposed slot exchange mechanism complements this view by introducing a structured peer interaction that reshapes agents’ action spaces. Rather than redefining the underlying game model, slot exchange provides a mechanism-level enhancement that allows agents to improve individual utility, expressed as comfort, while preserving convergence toward system-efficient outcomes. This positions the approach between cooperative and noncooperative paradigms: agents act autonomously and rationally, yet voluntarily engage in peer exchanges that lead to mutually beneficial and fair outcomes.
	
	\subsection{ Positioning and novelty of the proposed approach}
	
	In contrast to existing demand side management approaches that improve comfort primarily by regulating environmental variables such as temperature or illumination, this work improves comfort by coordinating consumption schedules directly through decentralized interaction. Unlike cooperative scheduling frameworks that rely on centralized optimization with ex post settlement, and unlike pricing-based or Stackelberg games that influence behavior indirectly, the proposed method introduces a lightweight slot exchange mechanism as an explicit coordination primitive.
	
	Slot exchange is not a local correction heuristic applied after optimization. Instead, it is integrated into a decentralized multi-agent optimization framework and enables agents to reduce discomfort through peer-level negotiation while respecting global inefficiency constraints. Furthermore, this work goes beyond most existing studies by systematically analyzing consumer behavior under varying levels of selfishness and altruism, and by evaluating fairness using variance and Gini coefficients.
	
	By explicitly linking decentralized optimization, peer coordination, comfort, fairness, and behavioral heterogeneity, the proposed framework fills a gap between cooperative demand side management, P2P flexibility exchange, and game-theoretic energy management, and provides a scalable and user-centric alternative to existing approaches.

	\begin{table}[h]
		\centering
		\caption{Notations}
		\label{tab:notations}
		\begin{tabular}{ll}
			\toprule
			\textbf{Symbol} & \textbf{Description} \\
			\midrule
			$A$ & Set of all agents in the system, $A = \{1, 2, \ldots, n\}$ \\
			$a$ & A single agent from the set $A$ \\
			$n$ & Total number of agents in the system \\
			$P_a$ & Set of all feasible plans for agent $a$ \\
			$\rho_{a,i}$ & $i$-th feasible plan for agent $a$ \\
			$\rho_{a,s}$ & Selected plan for agent $a$ \\
			$\rho_{a,p}$ & Preferred (most comfortable) plan for agent $a$ \\
			$d$ & Number of time slots in a plan \\
			$k$&Number of plans\\
			$s^a_i$ & Value of time slot $i$ in selected plan $\rho_{a,s}$ for agent $a$ \\
			$p^a_i$ & Value of time slot $i$ in preferred plan $\rho_{a,p}$ for agent $a$ \\
			$D_{a}$ & Discomfort cost of  agent $a$ \\
			$\overline{D}$ & Average discomfort cost across all agents \\
			$I$ & Inefficiency cost of the aggregated demand \\
			$g$ & Global demand plan, $g = \sum_{a=1}^{n} \rho_{a,s}$ \\
			$U$ & Unfairness metric (standard deviation of discomfort costs) \\
			$E$ & Total energy consumption in the system (constant) \\
			$\beta$ & Behaviour parameter ($0$ = selfish, $1$ = altruistic) \\
			$f_D(\cdot)$ & Discomfort evaluation function \\
			$f_I(\cdot)$ & Inefficiency cost evaluation function \\
			$C_{a}$& Comfort of agent $a$\\
			$G_{a}$& Comfort gain of agent $a$\\
			$R$&Slot exchange requests\\
			$T$&Iterations\\
			\bottomrule
		\end{tabular}
	\end{table}
	
	\section{Problem formulation}\label{PF}
	In this section, we define the key elements of the considered residential demand-side management scenario and formulate the main performance metrics of the proposed framework. The mathematical notations used throughout this paper are summarized in Table~\ref{tab:notations}.
	
	\subsection{Plan generation and representation}

	We consider a residential demand-side management problem modeled as a multi-agent system. Each energy consumer is represented by an autonomous software agent $a$, and the set of all agents is denoted by $A = \{1,2,\ldots,n\}$. Energy consumption is scheduled over $d$ discrete time slots within a day.
	
	Each agent $a \in A$ is associated with a finite set of energy consumption plans
	\[
	P_a = \{\rho_{a,1}, \rho_{a,2}, \ldots, \rho_{a,k}\},
	\]
	where each plan $\rho_{a,i} \in P_a$ is represented as a vector of length $d$. All plans satisfy a fixed total daily energy consumption constraint and differ only in their temporal distribution of energy across time slots.
	
	Among these plans, agent $a$ has a preferred plan $\rho_{a,p}$, which corresponds to the most comfortable energy consumption pattern and yields the minimum possible discomfort. During the optimization process, each agent selects exactly one plan from its plan set, referred to as the selected plan $\rho_{a,s}$. Formally,
	\begin{itemize}
		\item the preferred plan is $\rho_{a,p} = \{p_1^a, p_2^a, \ldots, p_d^a\}$,
		\item the selected plan is $\rho_{a,s} = \{s_1^a, s_2^a, \ldots, s_d^a\}$,
	\end{itemize}
	where $p_i^a$ and $s_i^a$ denote the energy consumption of agent $a$ in time slot $i$ under the preferred and selected plans, respectively.
	
	The selection of plans across all agents determines the aggregate demand profile of the system and directly affects system inefficiency, individual discomfort, and fairness. In this work, the objective is to identify a combination of selected plans that balances these competing criteria while satisfying energy conservation constraints.

	\subsection{Discomfort, Comfort, and Comfort Gain}
	
	An agent’s preference for a feasible plan is determined by the level of discomfort associated with deviations from its preferred energy consumption pattern. Let $\rho_{a,p}$ denote the preferred plan of agent $a \in A$, and let $\rho_{a,s}$ denote the selected plan. Each plan specifies energy consumption over $d$ discrete time slots. The values of energy consumption in time slot $i$ under the selected and preferred plans are denoted by $s_i^a$ and $p_i^a$, respectively.
	
	The discomfort cost of the selected plan $\rho_{a,s}$ for agent $a$ is defined as a function of the deviation between the selected and preferred plans:
	\begin{equation}
		D_{a} = f_D(\rho_{a,s}, \rho_{a,p}) 
		= \sqrt{\frac{1}{d} \sum_{i=1}^{d} \left( s_i^a - p_i^a \right)^2 } .
	\end{equation}
	This formulation quantifies discomfort using the root mean squared error (RMSE), capturing the average magnitude of deviations between the selected and preferred energy consumption schedules across all time slots.
	
	Based on the computed discomfort, consumer comfort is defined as a normalized measure reflecting the similarity between the selected and preferred plans. Specifically, the comfort of agent $a$ is given by
	\begin{equation}
		C_{a} = 1 - D_{a},
	\end{equation}
	where higher values of $C_{a} $ indicate greater comfort and correspond to deviations from the preferred energy consumption pattern.
	
	To evaluate the benefit of coordination, we define comfort gain as the improvement in comfort achieved by an agent relative to a baseline solution without coordination. The comfort gain is defined as
	\begin{equation}
		G_{a}  = C_{a} - C_a^{*},
	\end{equation}
	where  $C_a^{*}$ denote the comfort before optimization.

	\subsection{Inefficiency cost}The selected plans of all agents are aggregated to form a global plan $g = \sum_{a=1}^{n}\rho_{a,s}$. The global plan is the elemental sum  of selected plans of all agents, which is the total energy demand of all agents in a community over time. The inefficiency cost associated with the global plan is represented as \begin{equation}
		I =f_I(\sum_{a=1}^{n}\rho_{a,s}).
	\end{equation}
	Inefficiency cost is the difference between demand and supply of energy.
	A global response with lower cost is preferred over the one that has higher cost. 
	\subsection{Unfairness}
	In a multi-agent energy management system, fairness is essential to prevent some agents from consistently sacrificing their comfort for system-wide benefits. To capture this, unfairness is quantified as the dispersion of discomfort across agents and is defined as the standard deviation of individual discomfort costs.
	
	\begin{equation}
		U = \sqrt{ \frac{1}{n} \sum_{a \in A} (D_{a} - \overline{D})^{2} },
	\end{equation}
	where $U$ represents the system-wide unfairness, $D_a$ is the discomfort cost for agent $a$ and $\overline{D}$ is the average discomfort cost across all agents. A lower value of $U$ indicates a fairer distribution of costs among agents, whereas a higher value implies greater inequality. Within the proposed optimization framework, unfairness is minimized alongside inefficiency cost and individual discomfort to achieve a balanced system-wide solution.
	\subsection{Behavioral modeling of agents for demand-supply balance} 
	In a multi-agent energy management setting, the plan selected by an agent affects not only its own discomfort but also the aggregate demand of the system, and consequently the global inefficiency cost. Determining the optimal combination of plans across all agents therefore corresponds to a distributed combinatorial optimization problem, which is NP-hard \cite{shri-chu}. Each agent selects a plan $\rho_{a,s}$ by balancing its local discomfort against the impact of its decision on the global inefficiency cost. If agents act purely selfishly and prioritize only local comfort, the aggregate demand may significantly deviate from the available supply, leading to high inefficiency. Conversely, fully altruistic behavior, where agents prioritize system performance, may result in excessive local discomfort. To capture this trade-off, a behavioural parameter $\beta\epsilon[0,1]$ is introduced, which determines the relative importance assigned to local discomfort and global inefficiency during plan selection. A higher value of $\beta$ reflects more selfish behavior, while a lower value corresponds to more altruistic behavior. Accordingly, each agent selects its plan by minimizing a weighted combination of local discomfort and global inefficiency, as defined in the following equation \cite{farzam}.
	\begin{equation}\label{eq1}
		\rho_{a,s} = arg \overset{k}{\underset{s=1}{\min}} \Big( (1-\beta)\times f_I(g)+\beta \times f_D(\rho_{a,s})  \Big)
	\end{equation} 
	Equation \ref{eq1} shows that during the agent’s plan selection phase, information about global plan $g$ is required, which is not possible without coordination among agents. Making informed decisions is crucial for balancing energy demand and supply because when agents act independently and shift their load to off-peak hours to reduce costs, new peaks in energy demand can emerge.   
	\subsection{Coordinated plan adjustment and discomfort minimization}\label{sd}
	%
	%
	%
	%
	%

	After the decentralized plan selection phase, agents engage in a coordination process aimed at further improving individual comfort without affecting system efficiency. Each agent compares the time-slot values of its selected plan $\rho_{a,s}$ with those of its preferred plan $\rho_{a,p}$. For any time slot $i$ where $s_i^a \neq p_i^a$, the agent may request a slot exchange with another agent whose selected plan contains the desired value.
	
	Slot exchanges are performed pairwise between agents and are subject to the following constraints. First, the total energy consumption in the system must remain unchanged:
	\begin{equation}
		\label{con1}
		\sum_{a \in A} \sum_{i=1}^{d} s_i^a = E ,
	\end{equation}
	where $E$ denotes the fixed total energy consumption of the system. This constraint ensures that the aggregate demand profile remains balanced with the available supply.
	
	Second, each individual exchange must preserve the combined energy of the involved slots:
	\begin{equation}
		\label{con2}
		s_i^{a_1} + s_j^{a_2} = s_i^{a_1'} + s_j^{a_2'} ,
	\end{equation}
	where $s_i^{a_1}$ and $s_j^{a_2}$ denote the slot values of agents $a_1$ and $a_2$ before the exchange, and $s_i^{a_1'}$ and $s_j^{a_2'}$ denote the corresponding values after the exchange. This constraint guarantees that slot exchanges are energy-neutral and do not introduce local imbalances.
	
	The objective of the coordination process is to reduce overall discomfort by improving the alignment between selected and preferred plans across agents. Specifically, the coordination aims to minimize the average discomfort in the system, defined as
	\begin{equation}
		\min \overline{D} =
		\sqrt{
			\frac{1}{n \times d}
			\sum_{a \in A} \sum_{i=1}^{d}
			\left( s_i^a - p_i^a \right)^2
		}.
	\end{equation}
	
	Through such coordinated slot exchanges, agents can progressively reduce their individual discomfort and improve fairness, while strictly preserving the aggregate demand profile and the system-level inefficiency cost achieved during the initial optimization phase.

	\begin{figure*}
		\centering
		\includegraphics[width=1\linewidth]{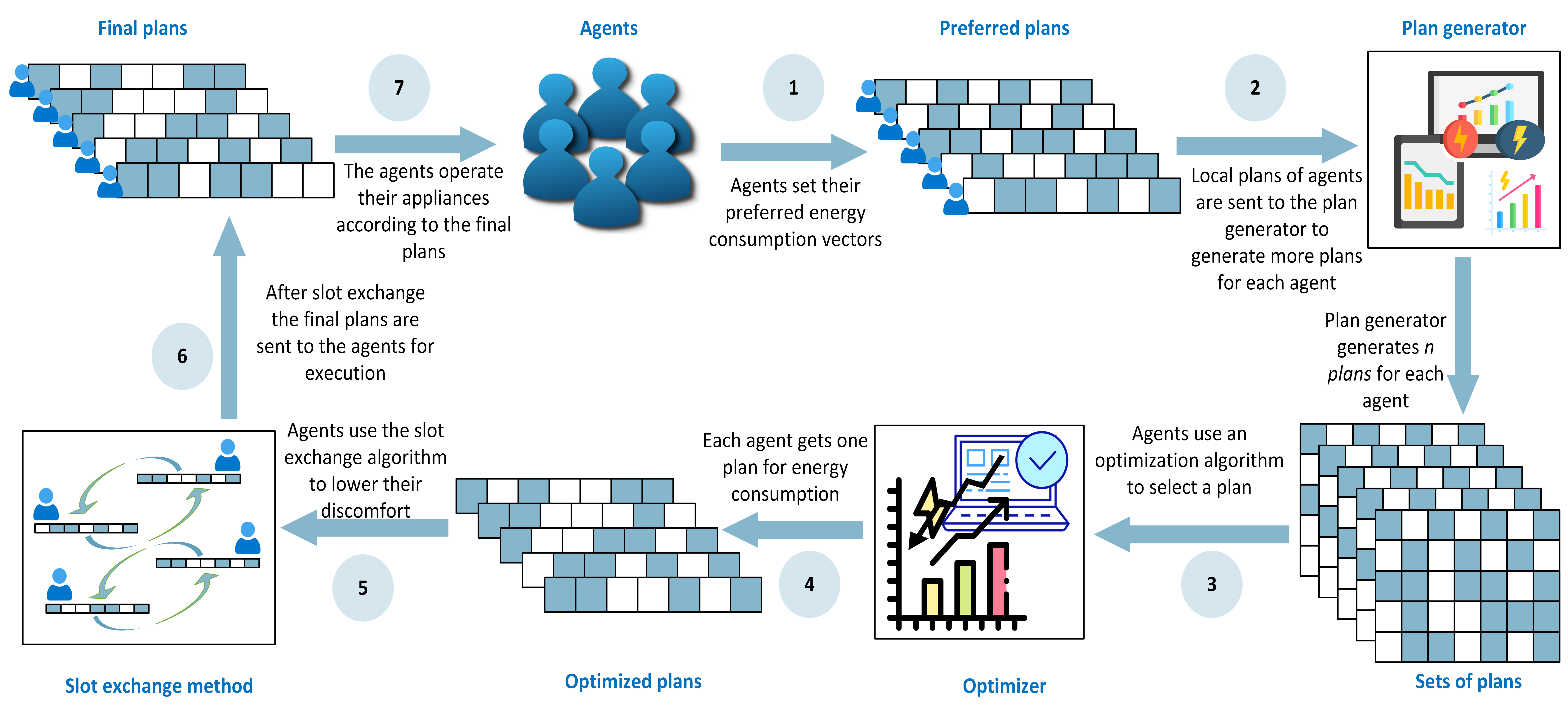}
		\caption{Proposed system model for energy optimization and cooperation of agents}
		\label{systemModel}
	\end{figure*}

	
	%
	%
	
	\section{Proposed Methodology}\label{PS}
	
	This section describes how the abstract problem formulation is instantiated in practice using a real-world dataset and decentralized coordination mechanisms. The overall system model is illustrated in \autoref{systemModel}.
	
	\subsection{Plan Generation and Flexibility Modeling}
	
	In this work, we rely on energy consumption plans that are already provided in the dataset and do not generate new plans during runtime. The plan-generation procedure follows the methodology introduced in \cite{farzam2}, where alternative plans are constructed by applying controlled permutation operations to a consumer’s baseline energy consumption schedule. Using pre-generated plans ensures reproducibility and allows us to focus on coordination and optimization rather than plan construction.
	
	For each agent, the dataset provides a baseline energy consumption schedule that corresponds to the preferred plan and is assigned a preference score of $1$, indicating maximum comfort. All alternative plans are derived from this baseline through permutation-based operations that redistribute energy consumption across time slots while preserving total daily energy consumption as well as the mean and standard deviation of the plan values.
	
	Specifically, the dataset includes three plan-generation schemes with increasing degrees of deviation from the preferred schedule:
	\begin{itemize}
		\item \textbf{SHUFFLE}: Three plans obtained by randomly permuting all time-slot values of the preferred plan, yielding the largest deviation and a preference score of $1/144$.
		\item \textbf{SWAP-15}: Three plans generated by randomly swapping values at 15 time-slot positions, with a preference score of $1/15$.
		\item \textbf{SWAP-30}: Three plans generated by randomly swapping values at 30 time-slot positions, with a preference score of $1/30$.
	\end{itemize}
	
	As a result, each agent is associated with a fixed set of $k=10$ plans: one preferred plan and nine alternative plans with varying degrees of deviation and preference scores in the range $[0,1]$. Consumer flexibility is implicitly encoded through the diversity of available plans, where larger deviations correspond to higher flexibility and lower comfort.
	
	The number of available plans $k$ directly impacts the computational complexity of the decentralized optimization process \cite{farzam2}. While a larger $k$ increases solution diversity and flexibility, it also increases communication and computation overhead. The choice of $k$, inherited from the dataset and prior work \cite{farzam}, represents a practical trade-off between expressiveness and scalability.

	\subsection{Decentralized Plan Selection via I-EPOS}
	
	To perform decentralized plan selection, the agents employ the Iterative Economic Planning and Optimized Selections (I-EPOS) algorithm \cite{epos-Energy, IEPOS}. I-EPOS is a collective learning mechanism designed for large-scale, privacy-preserving optimization in multi-agent systems. It organizes agents into a tree topology and iteratively coordinates decisions through bottom-up and top-down phases, enabling agents to balance local objectives with aggregated system-level information.
	
	Compared to alternative distributed optimization approaches such as consensus-based methods, centralized aggregators, or reinforcement learning techniques \cite{d1,d2,d3,d4}, I-EPOS offers several advantages that make it well suited for residential demand-side management. In particular, agents do not disclose their full energy consumption plans; instead, only aggregated information is exchanged, preserving consumer privacy. Moreover, the tree-based coordination structure significantly reduces communication overhead, allowing the system to scale to large agent populations.
	
	Using I-EPOS, each agent selects one plan from its available set by balancing local discomfort and system inefficiency, influenced by its behavioral parameter. The outcome of this stage is a system-feasible solution that achieves balanced electricity demand and supply. However, since plan selection is restricted to a finite set of alternatives, some agents may still experience sub-optimal comfort even after convergence, despite the global inefficiency cost being minimized.

	\subsection{Post-Optimization Slot Exchange Mechanism}
	
	After the decentralized plan selection phase converges, a slot exchange mechanism is applied as a post-optimization coordination layer to further improve individual comfort without modifying the  total energy consumption. This mechanism operates on the selected plans $\rho_{a,s}$ produced by the optimization phase and does not alter system-level inefficiency (respecting Equation \ref{con1}).
	
	 To facilitate efficient discovery of potential slot exchange partners, a lightweight \emph{match-making service} is introduced. The role of this service is limited to providing agents with information about which agents currently hold specific slot values. Importantly, the match-making service does not make decisions, enforce exchanges, or modify agent plans. All decisions regarding whether to initiate, accept, or reject a slot exchange are made autonomously by the agents themselves.
	
	Each agent compares its selected plan $\rho_{a,s}$ with its preferred plan $\rho_{a,p}$. For any time slot $i$ where $s_i^a \neq p_i^a$, the agent may request an exchange for the desired slot value. Rather than broadcasting requests to all agents, the agent queries the match-making service to identify candidate agents whose selected plans contain the desired value. This design significantly reduces communication overhead and avoids conflicting peer-to-peer negotiations.
	
	Although the match-making service introduces a centralized lookup component, the system remains decentralized in its decision-making and control. The service maintains only minimal slot-level information and does not have access to agents’ full plans, preferences, or discomfort values. As a result, privacy-sensitive information remains local to agents, and the coordination process scales linearly with the number of agents and slot queries.
	
	Once a potential exchange partner is identified, agents negotiate the exchange directly. An exchange is executed only if it satisfies energy neutrality constraints (\autoref{con1} and \autoref{con2}) and does not reduce the comfort of either participating agent.  For the requesting agent, the exchange is beneficial if the requested slot value matches its preferred slot value, i.e., replacing $s_i^a$ with $p_i^a$  reduces its discomfort. For the responding agent, an exchange is acceptable only if the slot being requested (selected slot) is not its preffered slot $s_i^a \neq p_i^a$. In this case, substituting the slot does not reduce its comfort. On the contrary, if the requested slot corresponds to the preferred slot of the responding agent, the exchange is immediately rejected in order to preserve local comfort priorities. Hence, the successfull slot exchange guarantees that the comfort of at least one agent improves while the comfort of the other agent does not decrease.
	
	To resolve conflicts when multiple agents request the same slot value, the match-making service processes slot exchange requests on a first-come–first-served basis. When a request is received by match-making service and a matching agent is identified, both the requesting agent and the matched agent are temporarily marked as unavailable. This prevents them from being considered for other exchange requests while the current negotiation is ongoing, thereby avoiding conflicting or inconsistent exchanges. 
	
	Once the two agents complete their local evaluation of the proposed exchange, they notify the match-making service of the outcome. If the exchange is successful, the corresponding slot values in the selected plans of both agents are updated. If the exchange is unsuccessful, the selected plans remain unchanged. In both cases, the availability status of the involved agents is reset, allowing them to participate in future exchange requests. This availability-based locking mechanism ensures consistent updates, eliminates race conditions, and allows conflict-free coordination without requiring centralized decision-making.

	No queuing mechanism is implemented for exchange requests. When a slot exchange occurs, the involved slot values are immediately updated, rendering any queued or pending requests invalid. This design choice avoids inconsistencies arising from outdated information and simplifies conflict handling. Through iterative exchanges, agents progressively reduce residual discomfort while strictly preserving the aggregate demand profile achieved during the optimization phase. The slot exchange process terminates when no successful slot exchange occurs during a complete iteration over all agents, indicating that no further comfort-improving exchanges are possible.

	\begin{algorithm}
		\caption{Multiagent energy optimization with slot exchange}
		
		\KwIn{
			$n$ $\leftarrow$ Number of agents\; 
			$k$ $\leftarrow$ Number of plans per agent\; 
			$d$ $\leftarrow$ Number of time slots per plan\; 
			$\beta$ $\leftarrow$ Weighting factor for inefficiency vs. discomfort cost\; 
			Preferred plans $\{\rho_{a,p} \;|\; \forall a \in A\}$
		}
		
		\KwOut{
			Final optimized plans for each agent after slot exchange
		}
		
		\ForEach{agent $a \in A$}{
			Generate $k$ feasible plans $P_a = \{\rho_{a,1}, ..., \rho_{a,k}\}$ using the preferred plan and flexibility value\;
			Compute local discomfort for each plan in $P_a$\;
		}
		
		Compute total inefficiency cost based on all agents' plans\;
		
		\Repeat{maximum iterations reached}{
			\ForEach{agent $a \in A$}{
				Evaluate objective function considering $\beta$-weighted trade-off between inefficiency cost and discomfort cost\;
				Exchange local information with neighbors to coordinate\;
				Update selected plan $\rho_{a,s}$ to achieve joint children/parents objective\;
			}
		}
		Update the matchmaking service about the selected plan of each agent\;
		\ForEach{agent $a \in A$}{
			Compare the value of slot $i$ in $\rho_{a,s}$ with the corresponding value of slot $i$ in $\rho_{a,p}$ for all $i \in {1, \dots, d}$\;
			Identify time slots where $s^{a}_{i} \neq p^{a}_{i}$\;
			\ForEach{such slot}{
				Request slot exchange via matchmaking service\;
				Matchmaking service provides info about agent $a^*$ whose slot value matches with the preferred slot value of a\;
				Matchmaking services updates the status of both agents to unavailable\;
				\If{$a^*$’s slot is not its preferred slot}{
					Negotiate slot exchange, ensuring energy balance\;
					Update plans of both agents by swapping slots\;
					Update the matchmaking service about the successful exchange\;
					Matchmaking sevice resets status of both agents to availablr\;
				}
			}
		}

		Repeat slot exchange until no further beneficial exchanges are possible\;
		
		\Return Final optimized plans for each agent\;
		
	\end{algorithm}

	 \subsection{Computational and Communication Complexity}

	\begin{table*}[t]
		\centering
		\caption{Computational and Communication Complexity Comparison}
		\label{tab:complexity}
		\begin{tabular}{lcc}
			\toprule
			\textbf{Algorithm Component} 
			& \textbf{Computational Complexity} 
			& \textbf{Communication Complexity} \\
			\midrule
			I-EPOS
			&$O(P \cdot T \log A)$
			& $O(T \log A)$ critical path \\
			Slot Exchange
			& $O(A \cdot d + A \cdot R)$
			& $O(R)$ \\
			Combined Approach
			& $O(P \cdot T \log A + A \cdot d + A \cdot R)$
			& $O(T \log A + R)$ \\
			\bottomrule
		\end{tabular}
	\end{table*}

	Table~\ref{tab:complexity} summarizes the computational and communication complexity of the proposed approach and its individual components, including I-EPOS and the slot exchange mechanism.
	
	As shown in Table~\ref{tab:complexity}, the I-EPOS algorithm exhibits low computational overhead per agent, since each agent evaluates a limited number of candidate plans over a fixed number of learning iterations. The communication complexity grows logarithmically with the number of agents due to the hierarchical aggregation structure employed by I-EPOS, which enables scalable decentralized coordination and has been shown to perform efficiently in large-scale settings.
	
	The slot exchange mechanism introduces additional computational and communication costs that depend on both the number of agents $A$ and the number of time slots $d$. The dominant computational cost of this phase arises from agents comparing assigned slots with preffered slots $O(A\cdot d)$ and processing slot exchange requests $R$ by comparing assigned slot values across agents for a given time slot $O(A \cdot R)$. This results in a computational complexity of $O(A\cdot d+A \cdot R)$, where $R$ denotes the total number of slot exchange requests. Communication during this phase scales linearly with the number of mismatching slots, as each exchange attempt requires only a constant number of message exchanges between agents.
	
	When both phases are combined, the total computational complexity is given by the sum of the I-EPOS plan selection cost and the slot exchange overhead. As indicated in Table~\ref{tab:complexity}, this yields an overall computational complexity of $O(P \cdot T \log A +A\cdot d+ A \cdot R)$. Where $P$ represents plans and $T$ iterations. In the worst case, where all time slots mismatch and $R = A \cdot d$, this complexity becomes $O(P \cdot T + A^2 \cdot S)$. However, this represents a conservative upper bound, as in the proposed energy management scenario, only a subset of slots requires reallocation (demand is already optimized using IEPOS), and the number of exchange requests remains significantly lower than the theoretical maximum. Hence, the computational cost will remain linear. 
	
	Similarly, the total communication complexity combines the logarithmic communication cost of the I-EPOS phase with the linear communication cost of the slot exchange mechanism, resulting in $O(T  \log A+ R)$. In the worst case, this simplifies to $O(T  \log A+ A \cdot S)$, while remaining linear in both the number of agents and the number of time slots.
	
	Overall, the results in Table~\ref{tab:complexity} highlight a trade-off between coordination overhead and solution refinement. While the slot exchange phase increases computational effort due to additional coordination among agents, it enables fine-grained improvements in individual agent satisfaction without compromising the decentralized and scalable nature of I-EPOS. This balance makes the combined approach well suited for large-scale energy management applications where solution quality and adaptability are critical.

	\begin{figure*}
		\vspace{-6em}
		\centering
		\includegraphics[width=0.9\linewidth]{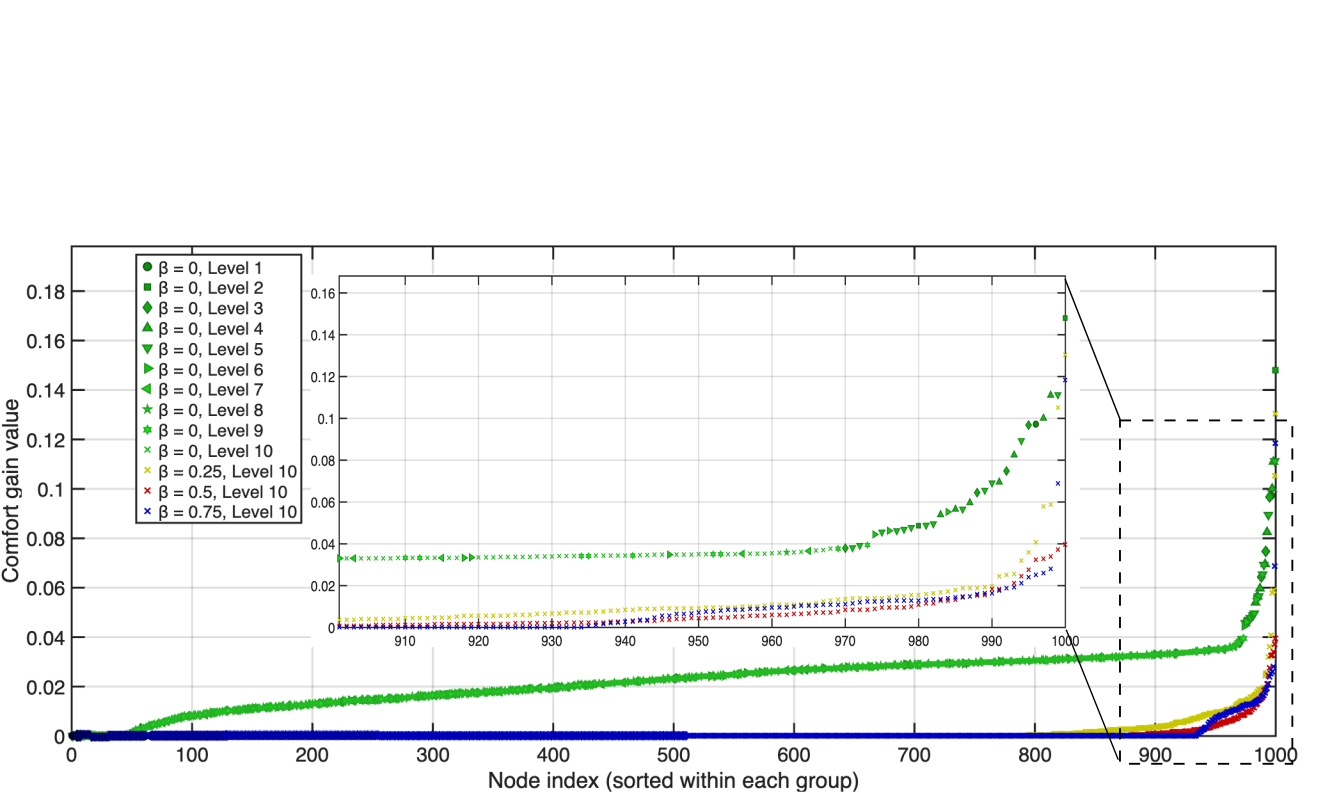}
		\caption{Comfort gain of agents under different $\beta$ values }
		\label{2}
	\end{figure*}
	\section{Experimental Evaluation} \label{SR}
	
	\subsection{Simulation environment}
	The energy management scenario in this study uses a dataset derived by disaggregating simulated zonal power transmission data for the Pacific Northwest \cite{dataset}. The detailed power consumption profiles for 1,000 residential consumers are analysed, each modeled as an agent. The energy consumption data is recorded at five-minute intervals over a 12-hour period. Consequently, each energy plan consists of $d=144$ time slots.
	
	For each agent, k=10 feasible energy plans are generated. To evaluate the proposed multi-agent coordination mechanism, a series of simulations is conducted using this dataset. For each simulation, the process is run over 50 iterations. A total of 10 simulation runs are performed to assess the consistency and robustness of the results. 
	
	The implementation of the proposed framework, including the I-EPOS optimization and the slot exchange mechanism, is publicly available in an open-source GitHub repository. The repository includes all software dependencies, and configuration files required to reproduce the reported results and figures. Detailed instructions for running the experiments and regenerating the plots are provided in the repository documentation.
	
	
	\subsection{Effect of behavioral parameter $\beta$ and slot exchange}
	The primary baseline in this work is the standard I-EPOS optimization without the proposed slot exchange mechanism. Specifically, the proposed approach (I-EPOS with slot exchange) is evaluated against the I-EPOS-only baseline under identical conditions, and this comparison is repeated for a range of $\beta$ values to reflect changing consumer behaviors. As shown in \autoref{2}, the comfort gain is computed by comparing the comfort achieved by agents under I-EPOS alone with the comfort achieved after applying the slot exchange mechanism. The results demonstrate that, across all tested $\beta$ values, the proposed coordination mechanism consistently improves user comfort without increasing inefficiency, while the variation in $\beta$ highlights how these improvements interact with different behavioral tendencies of consumers.
	
	In I-EPOS, agents are organized hierarchically in a binary tree with nine levels (L1 to L9). Due to the nature of the binary structure, the majority of agents are located at the lowest level (L9). The system uses the weighting factor $\beta$ to adjust the balance between selfish (discomfort cost) and cooperative (system inefficiency) objectives. With higher $\beta$ values agents tend to prioritize their individual comfort, whereas smaller values promote agents' collective behavior focused on reducing system-level inefficiency.
	
	\autoref{2} shows the sorted comfort gain for all agents under four scenarios: $\beta = 0$, 0.25, 0.5, and 0.75. The figure uses a combination of color gradients and marker styles to encode the hierarchical levels of the tree. Agents nearer the root (L1) are shown with darker hues and characteristic markers, whereas lower-level agents (L9) are represented with progressively lighter tones and distinct marker shapes.
	
	When $\beta$ is set to zero, all agents act altruistically during plan selection, focusing entirely on minimizing the inefficiency cost rather than their individual preferences. This leaves greater room for agents to improve their comfort through slot exchange, resulting in the highest overall comfort gain and the widest participation across all levels. In this scenario, slot exchanges occur at all levels, from L1 to L9, leading to a more equitable distribution of comfort gain. As $\beta$ increases, agents behave more selfishly by selecting plans that more closely align with their preferred consumption patterns during the I-EPOS optimization. Consequently, the need and opportunity for slot exchange diminish, since more agents are likely to already obtain their desired slots, and only a small number of agents who did not obtain their preferred slots participate in slot exchanges. The results show that with high $\beta$ values (e.g., 0.75), agents at the higher levels benefit the most, achieving near-optimal comfort, while lower-level agents, especially those at L9, have fewer opportunities to improve. This behavior arises from the hierarchical coordination mechanism of I-EPOS, where agents located closer to the root of the aggregation tree converge earlier and have the opportunity to select the plans that are closer to their comfort optima.
	
	Zero comfort gain indicates agents that did not participate in the slot exchange phase, either because their initially assigned plans already aligned with their comfort preferences or because no feasible exchange match was identified. The \autoref{2} depicts that the proposed slot exchange mechanism is effective and more useful when agents show altruism and prioritize inefficiency cost over discomfort cost. 
	\begin{figure}
		\centering
		\includegraphics[width=1\linewidth]{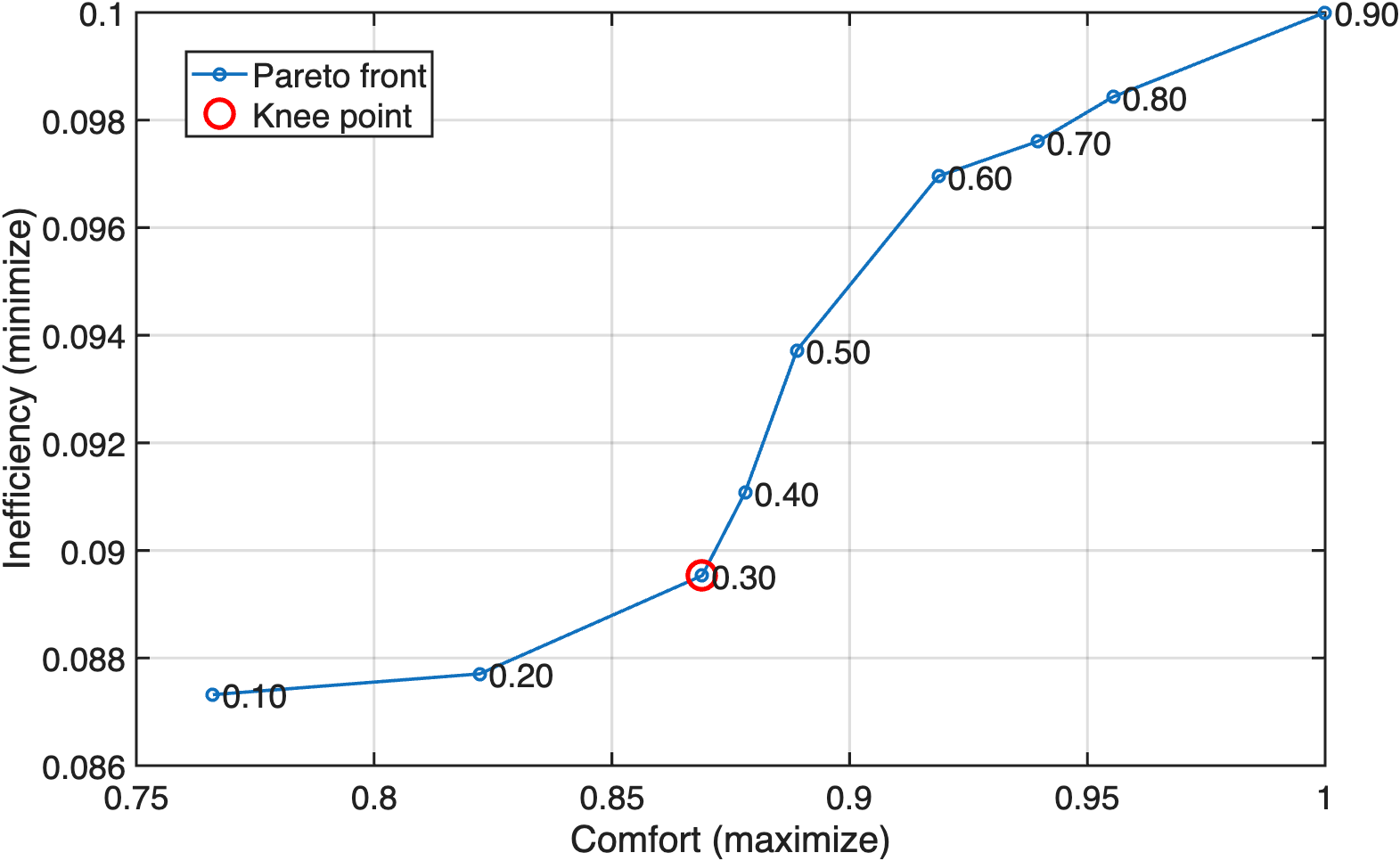}
		\caption{Pareto front between consumer comfort and system inefficiency, highlighting the knee point at $\beta = 0.3$.}
		\label{fig:pretoknee}
	\end{figure}
	
	\autoref{fig:pretoknee} illustrates the Pareto front between consumer comfort and system inefficiency, for different values of the behavioral parameter $\beta$. Each point on the curve corresponds to the outcome obtained for a specific $\beta$. The Pareto front exhibits a clear knee at $\beta$ = 0.3, highlighted in the figure. Up to this point, increasing $\beta$ yields simultaneous and meaningful improvements in comfort with only a marginal increase in system inefficiency. Beyond the knee, further increases in $\beta$ result in diminishing returns: comfort improvements become progressively smaller, while inefficiency increases more rapidly. This change in curvature indicates that $\beta$ = 0.3 represents a balanced operating point where the trade-off between individual comfort and system performance is most favorable. From a behavioral modeling perspective, this knee point can be interpreted as the emergent compromise between self-interest and collective efficiency. Rather than treating $\beta$ as an arbitrary tuning parameter, the Pareto knee provides a principled way to select $\beta$ based on observed system-level outcomes. 
	
	In real-world scenarios, $\beta$ could be calibrated using historical participation data, observed user responses to incentives, or empirical measurements of willingness to shift consumption in response to price or coordination signals.
	
	\begin{figure}
		\centering
		\includegraphics[width=1\linewidth]{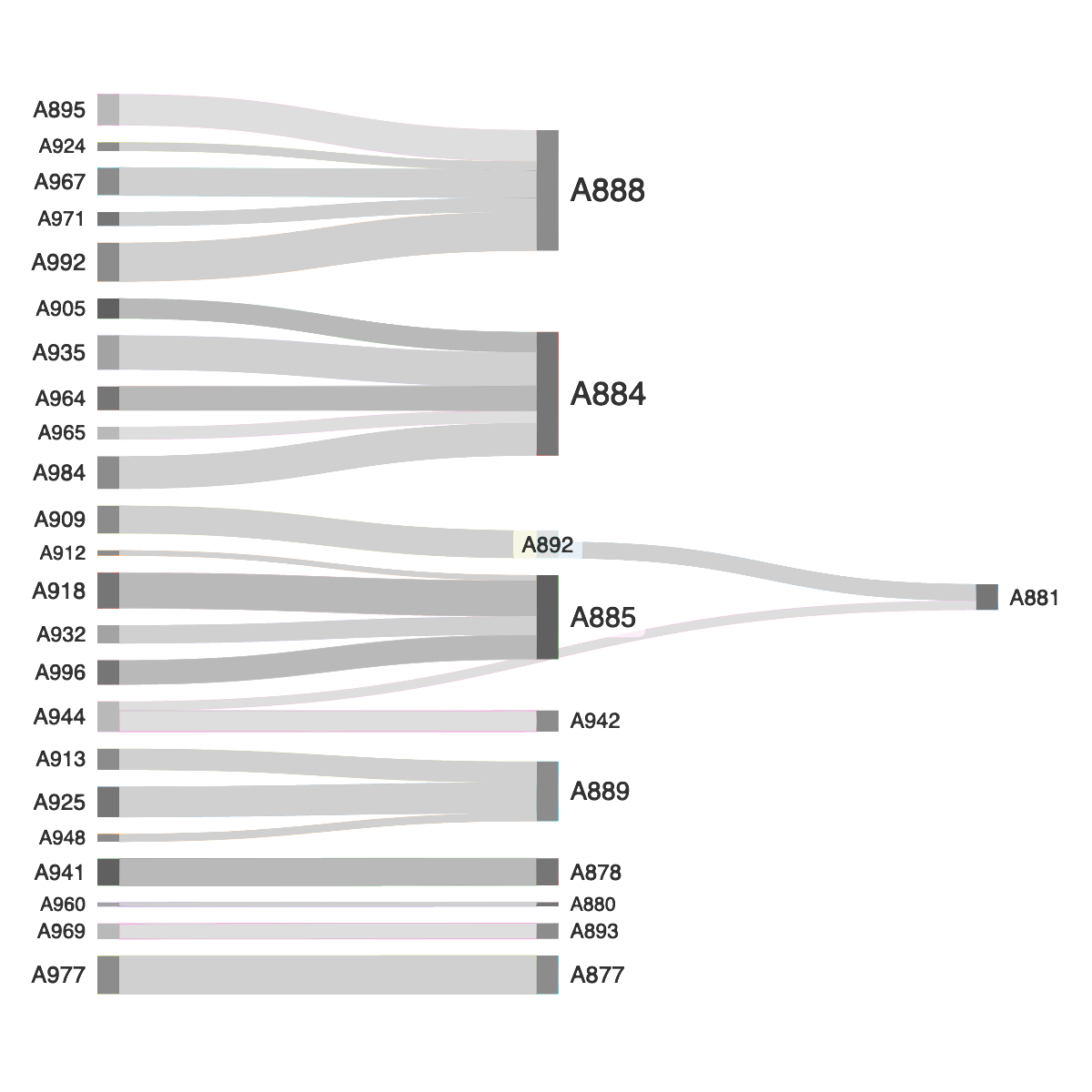}
		\caption{A subset of agents interactions for slot exchange}
		\label{sanky}
	\end{figure}
	
	\autoref{sanky} presents the Sankey diagram depicting agent interactions during the slot exchange mechanism in the proposed demand side management system. This figure presents a subset of agents participating in the slot exchange phase to illustrate how individual comfort is improved while preserving overall system balance. The example is taken from the scenario where $\beta$ value is 0 and more agents take part in the slot exchange mechanism. In the diagram, agents positioned on the left side act as slot exchange request initiators, while those on the right side accept incoming requests and participate in the exchange.
	
	Each connection represents a successful slot transfer between two agents, with the width of the connection weighted by the resulting comfort gain. Thicker connections indicate larger improvements in comfort and more substantial exchanges. Notably, certain agents ($30-40\%$), such as A892, perform dual roles by both initiating requests and accepting slots from others. This dual participation underscores the pivotal role of such agents. The diagram also highlights hierarchical groupings of agents and localized interactions, such as those involving A888, which interacts with multiple agents, such as A895 and A924, forming a cluster of exchanges. Similarly, A885 connects with both A892 and A881, facilitating downstream exchanges. From this plot, we learn that in the proposed system, agents can act both as request initiators and as acceptors, and some even participate simultaneously in both roles.

	\begin{figure}
		\centering
		\includegraphics[width=1 \linewidth]{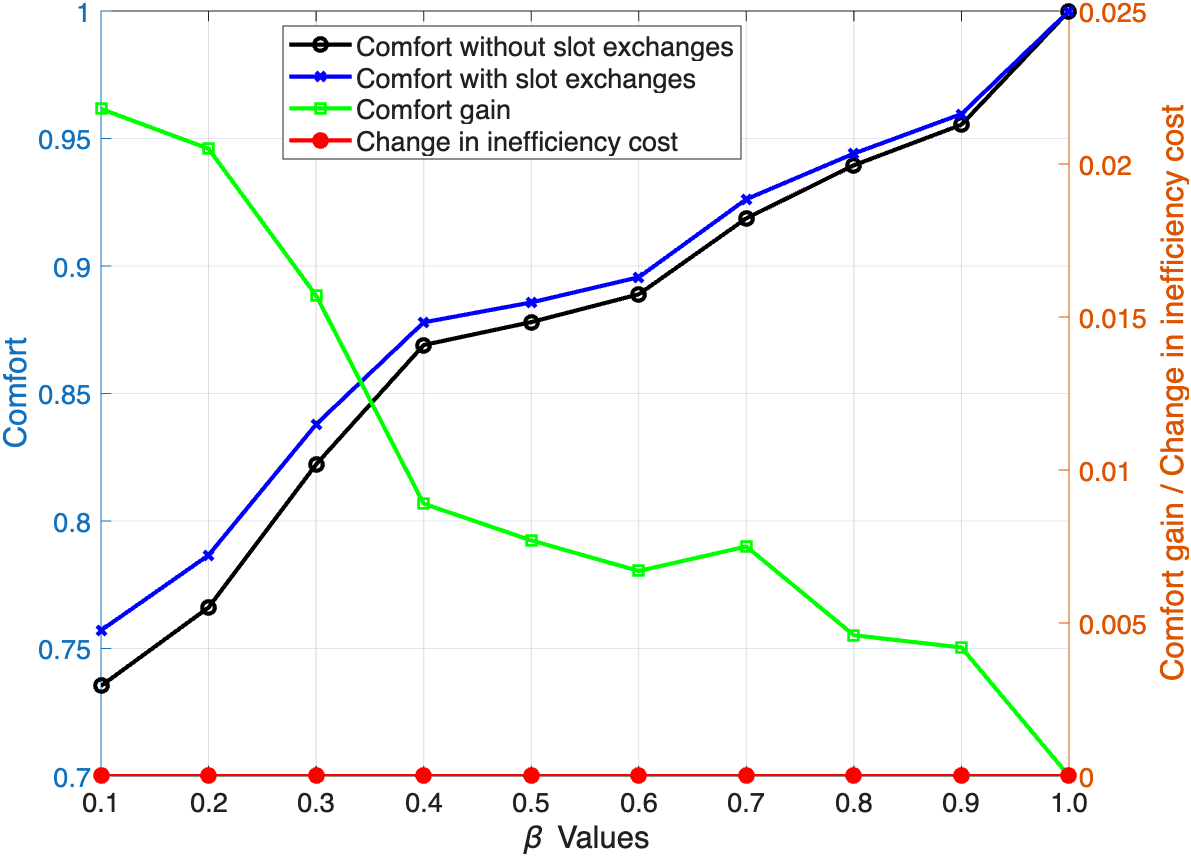}
		\caption{Effect of slot exchange method on average comfort of agents. Higher $\beta$ value means lower comfort gain}
		\label{fig:comfortcomparison}
	\end{figure}
	
	\autoref{fig:comfortcomparison} illustrates the effect of the slot exchange mechanism on agents’ comfort across different $\beta$ values in the I-EPOS optimization framework. The results highlight a key insight: as the $\beta$ value increases, more agents achieve their maximum comfort by selfishly selecting their preferred slots during the initial optimization phase. This behavior reduces the need for additional improvements through the slot exchange mechanism. Consequently, the comfort gains resulting from the slot exchange process decrease with higher $\beta$ values because fewer agents seek to further enhance their satisfaction, which limits the number of compatible slot matches and reduces the frequency of exchanges. 
	
	However, the slot exchange algorithm demonstrates significant effectiveness, particularly at lower $\beta$ values. In these scenarios, a larger proportion of agents, more than 80\%, require additional optimization to improve their comfort, and the algorithm facilitates extensive exchanges, leading to substantial overall gains. It is important to note, though, that not all agents succeed in increasing their comfort, since it is not guaranteed that every agent finds a compatible match. Agents rely on the matchmaking service that shares information about available slots for exchange. As a result, around 60\% of agents successfully find matching slots and benefit from the mechanism. It is also important to note that the enhancement of comfort with the slot exchange mechanism does not affect the inefficiency cost. The change in inefficiency cost remains zero. 
	
\begin{figure}
		\centering
		\includegraphics[width=0.9\linewidth]{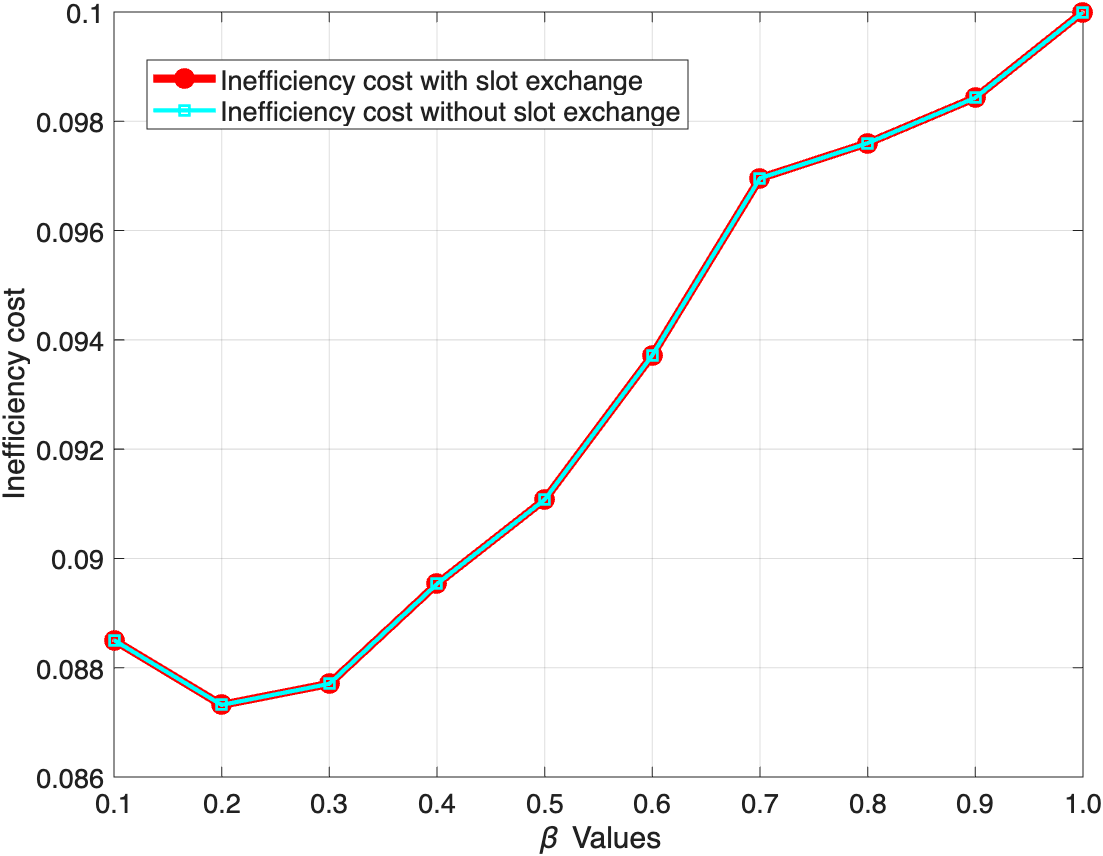}
			\caption{Inefficiency cost increases with the increase in $\beta$ value, and also there is no change in inefficiency cost after using slot exchange method.}
		\label{inefficiency}
	\end{figure}

	Moreover, even at higher $\beta$ values, the slot exchange mechanism maintains its utility by ensuring that the remaining agents who have not achieved their maximum comfort can still benefit from this collaborative comfort optimization. This highlights the versatility and robustness of the proposed algorithm, as it adapts to varying degrees of selfish behavior ($\beta$ values) while consistently enhancing comfort where possible. Moreover, the success rate of exchange requests remains high because agents only advertise slots they do not prefer in their selected plan, increasing the likelihood of feasible matches. Moreover, \autoref{inefficiency} depicts that higher $\beta$ values also lead to an increase in inefficiency cost, which is not ideal for the grid side. This reflects the trade-off between individual comfort and overall system efficiency, as agents prioritizing their own preferences at higher $\beta$ values impose a higher burden on the grid. The slot exchange mechanism, while primarily aimed at improving comfort, does not affect the inefficiency cost, ensuring that the system maintains a balance between comfort enhancement and grid stability.

	\subsection{Scalability and comfort evaluation}
	
	\begin{figure}
		\centering
		\includegraphics[width=1\linewidth]{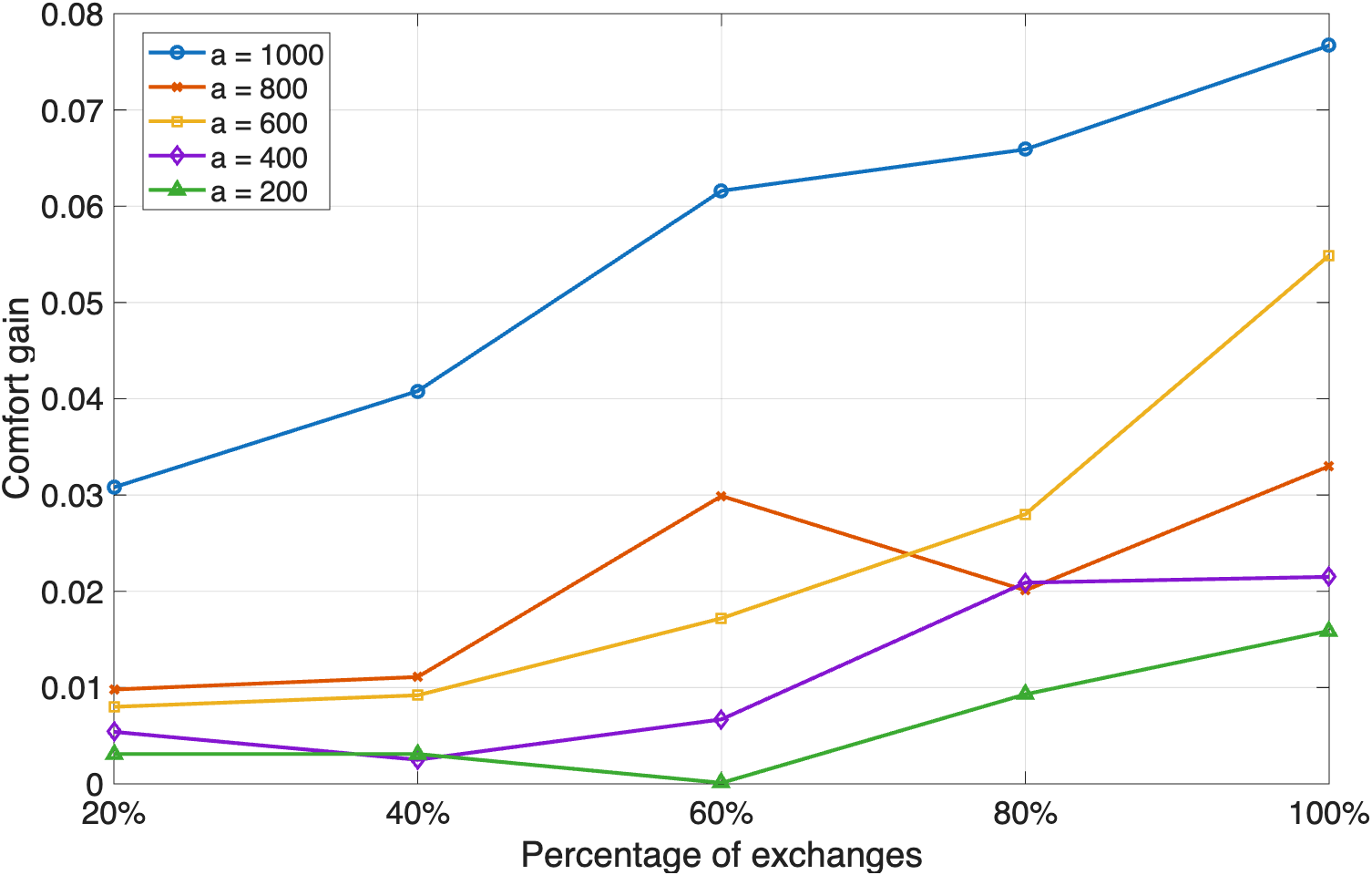}
		\caption{The average comfort gain under different population sizes}
		\label{fig:comfortgain}
	\end{figure}

	\begin{figure}
		\centering
		\includegraphics[width=1\linewidth]{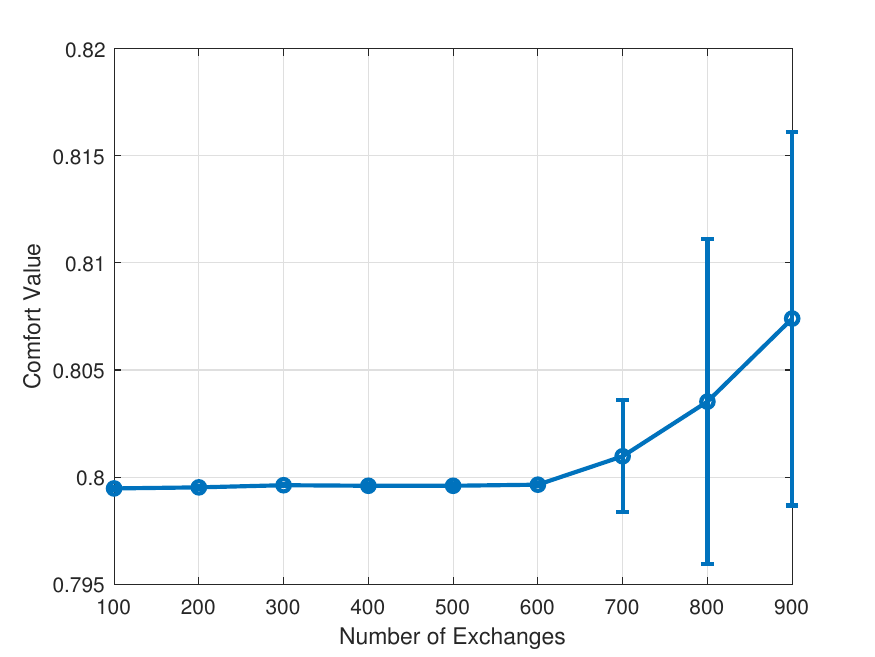}
		\caption{The effect of slot exchanges on average comfort value of agents}
		\label{fig:comfortvalueevolutionwithincreasingexchangesandlevels}
	\end{figure}
	Furthermore, the effectiveness of the slot exchange mechanism is evaluated under different population sizes, as illustrated in \autoref{fig:comfortgain}, to assess the scalability of the proposed approach and to examine whether comfort improvements persist as the number of agents increases. Population sizes of 200, 400, 600, 800, and 1,000 agents are considered, and the average comfort gain is computed for each case. To analyze the impact of participation density, each population is further divided into randomly selected subsets comprising 20\%, 40\%, 60\%, 80\%, and 100\% of the agents. For each subset, slot exchanges are performed among the selected agents and the resulting average comfort gain is evaluated. This binning approach enables an assessment of how increasing the number of potential exchange participants influences achievable comfort improvements.
	
	The results indicate that increasing both the overall population size and the proportion of participating agents generally expands the slot exchange space, leading to more feasible exchange opportunities and higher average comfort gains. For example, in the largest population of 1,000 agents, increasing the participating subset from 40\% to 60\% raises the average comfort gain from 0.04 to 0.06, while a further increase to 80\% results in only a marginal improvement to 0.065. This demonstrates that although additional exchanges tend to improve comfort, the relationship is not linear. In some cases, increasing the number of participating agents does not lead to higher comfort gains. For instance, for a population of 800 agents, expanding the participating subset from 60\% to 80\% results in a slight decrease in average comfort gain. This behavior can be attributed to the inclusion of additional agents with limited flexibility or low-comfort plans, which may not find compatible exchange partners and therefore do not contribute effectively to overall improvement.
	
	In addition to analyzing population size, the impact of the number of slot exchanges on agent comfort is explored, as illustrated in \autoref{fig:comfortvalueevolutionwithincreasingexchangesandlevels}. To evaluate this, randomized subsets of slot exchanges from all 1000 agents are created in increasing sizes from 100 to 900 exchanges. For each exchange size (100, 200, ..., 900), five independent sets are randomly selected, and the average comfort value is computed for each set.
	
	The results reveal that the effectiveness of slot exchanges is not uniform across different numbers of slot exchanges. While some subsets lead to substantial comfort improvements, others yield marginal gains. This disparity suggests that not all slot exchanges contribute equally to overall comfort enhancement some have a more significant impact due to the comfort gap they resolve for the involved agents.
	
	Up to an exchange size of 600, the variation in average comfort values across sets remains relatively minor, typically fluctuating between 0.791 and 0.800. However, starting from 700 exchanges, the variation becomes more noticeable. For instance, at 700 exchanges, average comfort values range from 0.797 to 0.804, while at 900 exchanges, they vary more widely from 0.798 to 0.819. This increasing spread indicates that larger sets of exchanges introduce more diversity in outcomes, reflecting the system's dynamic and context-dependent nature.
	
	The variation in comfort outcomes is a direct consequence of how comfort is computed and how individual slot exchanges affect the underlying deviation. Comfort is quantified using the RMSE between an agent’s assigned load profile and its preferred profile. If the exchanged slot have larger deviation between assigned and preferred energy, the exchange produces a substantial reduction in the squared error term of the RMSE, leading to a noticeable increase in comfort. Conversely, if the exchanged slot initially exhibits only a small deviation, the resulting change in the RMSE is limited, even though an exchange has taken place. As a result, different slot exchanges yield different impacts on overall comfort depending on the magnitude of the slot-level deviation they address. Therefore, the observed variability in comfort is not arbitrary but is determined by the deviation magnitude of the exchanged slots.
	
	\subsection{Fairness evaluation}
	\begin{figure}
		\centering
		\includegraphics[width=1\linewidth]{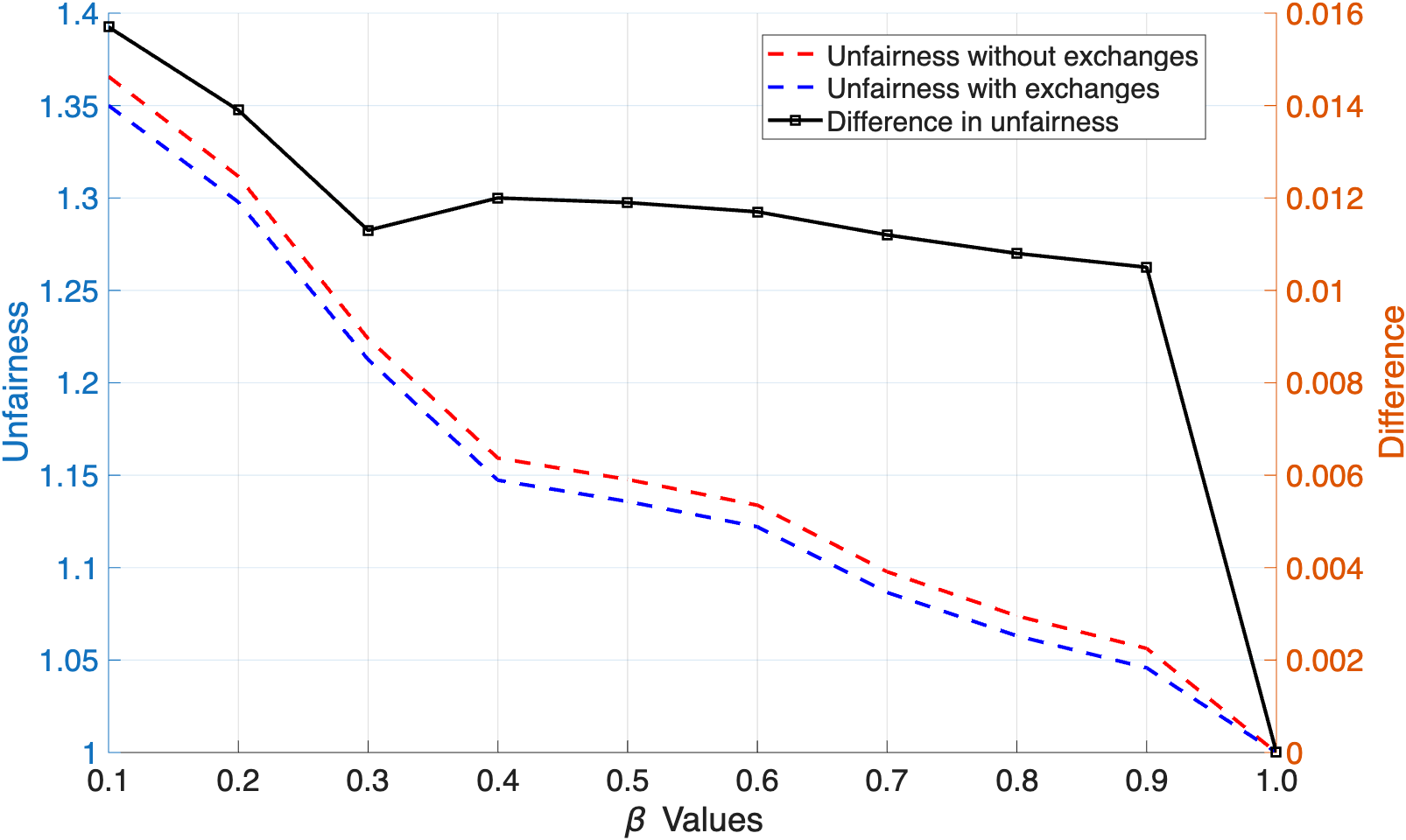}
		\caption{The effect of slot exchange mechanism on unfairness in the system}
		\label{fig:unfairness}
	\end{figure}
	\autoref{fig:unfairness} illustrates the effect of the proposed slot exchange mechanism on system unfairness across different $\beta$ values in the I-EPOS framework. As $\beta$ increases, agents behave more selfishly and are more likely to select their preferred slots during the initial optimization phase. This leads to a more uniform distribution of comfort, as more agents are individually satisfied, thereby reducing overall unfairness in the system.
	
	Crucially, the introduction of the slot exchange mechanism further improves fairness, even after the initial allocation. By enabling agents to reallocate non-preferred slots with others who find them more desirable, the mechanism compensates for disparities that may have resulted from the optimization process especially at lower $\beta$ values where agents act more altruistically and sacrifice personal comfort for global efficiency.
	
	At lower $\beta$ values, some agents may end up with significantly less comfort due to their cooperative behavior. However, the slot exchange process mitigates these disparities, allowing comfort levels to be redistributed more equitably. As a result, the system becomes fairer, and no group of agents is systematically disadvantaged.
	
	The results in \autoref{fig:unfairness} clearly demonstrate that both increasing $\beta$ and applying the slot exchange mechanism contribute to reducing unfairness. While higher $\beta$ values inherently support fairer outcomes by granting more agents their preferred allocations, the slot exchange mechanism ensures that residual unfairness is addressed through collaborative reallocation. This reinforces the effectiveness of the proposed method not only in enhancing overall comfort but also in promoting equity and fairness among agents under various system conditions.
	without increasing inefficiency.
	\begin{figure}
		\centering
		\includegraphics[width=1\linewidth]{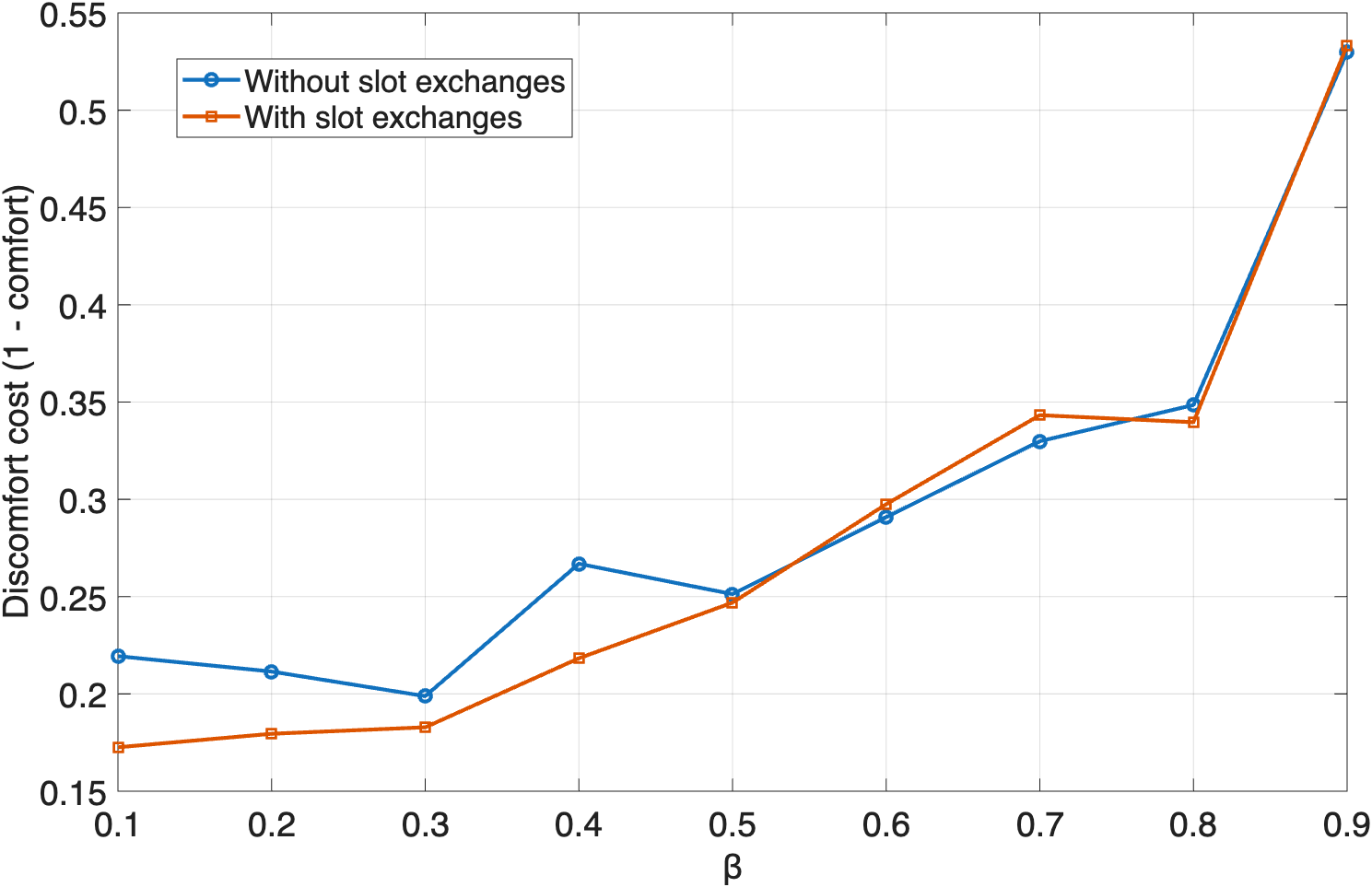}
			\caption{Gini coefficient of comfort distribution across agents under varying $\beta$ values, comparing I-EPOS with and without the proposed slot exchange mechanism.}
		\label{fig:ginni}
	\end{figure}
	To complement the standard deviation–based unfairness analysis, we additionally evaluate the distribution of discomfort using the Gini coefficient, which captures structural inequality in comfort outcomes across agents. While the standard deviation reflects dispersion around the mean, it does not distinguish between random variability and systematic concentration of discomfort among specific agents. In contrast, the Gini coefficient explicitly measures inequality in the distribution and is sensitive to whether discomfort is unequally distributed between agents.
	
	 \autoref{fig:ginni} compares the Gini coefficient of discomfort with and without the slot exchange mechanism for different values of the behavioral parameter $\beta$. Across most $\beta$ values, the inclusion of slot exchange reduces the Gini coefficient, indicating a more equitable distribution of discomfort among agents. This suggests that slot exchanges not only reduce average discomfort but also mitigate structural inequalities by redistributing unfavorable slots away from systematically disadvantaged agents.
	
	As $\beta$ increases, the Gini coefficient rises in both cases, reflecting the increasing concentration of discomfort under more self-interested behavior. However, the consistently lower Gini values observed with slot exchanges demonstrate that the proposed mechanism alleviates inequality even when agents prioritize individual comfort. These results confirm that slot exchange improves fairness in a structural sense, complementing the variance-based analysis and providing a more complete characterization of equity outcomes.
	
	\section{Limitations and Future Work}\label{flim}
	
	The proposed framework adopts several modeling abstractions to enable a clear analysis of comfort-aware coordination in residential demand-side management. In particular, the slot exchange mechanism operates at an aggregate slot level and assumes that slot values are directly swappable between agents. This abstraction ensures energy neutrality and preserves the aggregate demand profile, allowing us to isolate the effect of coordination on comfort and system inefficiency. However, real-world energy consumption is driven by heterogeneous appliances with operational constraints such as indivisible cycles and non-shiftable loads, which are not explicitly modeled in this study. In addition, plan sets are pre-generated and remain fixed during the optimization phase, limiting agents’ ability to dynamically adapt plan values during coordination. Finally, the match-making service introduces a lightweight centralized lookup component, which, while non-decisional and privacy-preserving, may require distributed implementations for large-scale deployments.
	
	Future work will focus on extending the proposed approach toward more realistic and deployable settings. This includes incorporating appliance-level datasets that explicitly distinguish between shiftable and non-shiftable loads, enabling feasibility checks for slot exchanges at the appliance level. We also plan to integrate the slot exchange mechanism more tightly within the I-EPOS coordination process, allowing plan values to be dynamically adjusted during optimization rather than only as a post-processing step. Additional directions include sensitivity analyses on flexibility distributions, plan set sizes, and coordination depth, as well as distributed implementations of the match-making service to further enhance scalability, robustness, and privacy.

	\section{Conclusion} \label{con}
	This paper presents a decentralized energy management approach that integrates a slot exchange mechanism to enhance individual comfort of consumers without affecting system inefficiency cost. The results show that while agents act more selfishly, achieving their preferred slots and reducing the need for further optimization, the proposed slot exchange mechanism remains crucial. It consistently improves overall comfort, even in less cooperative settings, by enabling agents to engage in mutually beneficial slot exchanges. 
	
	Moreover, the results highlight that fairness in comfort distribution improves through the slot exchange process. As population size increases, the system exhibits higher comfort gains due to more opportunities for matching agents with complementary needs. However, the relationship is not strictly linear, as some newly added agents may not always find suitable exchange partners, underscoring the dynamic nature of decentralized coordination.
	
	The key takeaway is that the proposed slot exchange mechanism enhances comfort in a scalable and adaptive manner.  It provides a robust foundation for real-world applications in utility-led demand side management programs, route planning for vahicles and drones,vehicle charging schedule optimization, etc. By embedding this mechanism into decentralized decision-making systems, system designers and practitioners can improve scheduling decisions, reduce consumer dissatisfaction, and promote fairer outcomes across diverse population sizes and behavioral profiles.

	\section*{Declarations}
	\subsection{Ethics approval and consent to participate}
	Not applicable
	\subsection{Consent for publication}
	Not applicable
	\subsection*{Availability of data and material}
	All experiments were implemented in Java using the JADE multi-agent framework and developed within the Eclipse IDE. The source code developed and used in this study is openly available at
	\url{https://github.com/rbyakhalid/DSM}.
	The dataset supporting the findings of this study is publicly available at Figshare:
	\url{https://figshare.com/articles/dataset/Optimality_Benchmark_for_Combinatorial_Optimization/7803005}.
	The repository includes a detailed description of software dependencies, execution instructions, and scripts for reproducing all figures and tables reported in this paper. 
	
	\subsection*{Competing interests}
	The author declares that there are no competing interests.
	
	\subsection*{Funding}
	Not applicable.
	
	\subsection*{Authors' contributions}
	Rabiya Khalid conceived and implemented the research, developed the proposed energy management system, performed the experiments, and prepared the manuscript under the supervision of Evangelos Pournaras. Evangelos Pournaras supervised the work, provided feedback on the research ideas and experimental design, critically reviewed and edited the manuscript, and contributed to shaping and refining the study.
	
	\subsection*{Acknowledgements}
	This work is supported by a UKRI Future Leaders Fellowship (MR/W009560/1): ‘Digitally Assisted Collective Governance of Smart City Commons–ARTIO’.
	\section*{Abbreviations}
	
	\begin{table}[h!]
		\centering
		\begin{tabular}{ll}
			\hline
			\textbf{Abbreviation} & \textbf{Meaning} \\
			
			MASs & Multi-Agent Systems \\
			DMA-EMS & Decentralized Multi-Agent Energy Management System \\
			DR & Demand Response \\
			IGWO & Improved Grey Wolf Optimization \\
			AI&Artificial Intelligence\\
			GA & Genetic Algorithm \\
			PSO & Particle Swarm Optimization \\
			BA & Bat Algorithm \\
			CFD & Computational Fluid Dynamics \\
			BPNN & Back-Propagation Neural Network \\
			AMOPSO-GWO & Adaptive Multi-Objective PSO--Grey Wolf Optimization \\
			ENGO & Enhanced Northern Goshawk Optimization \\
			I-EPOS & Iterative Economic Planning and Optimized Selections \\
			\hline
		\end{tabular}
		
	\end{table}

	\onecolumn
	\appendix

	\begin{figure}[H]
		\centering
		\includegraphics[width=1\linewidth]{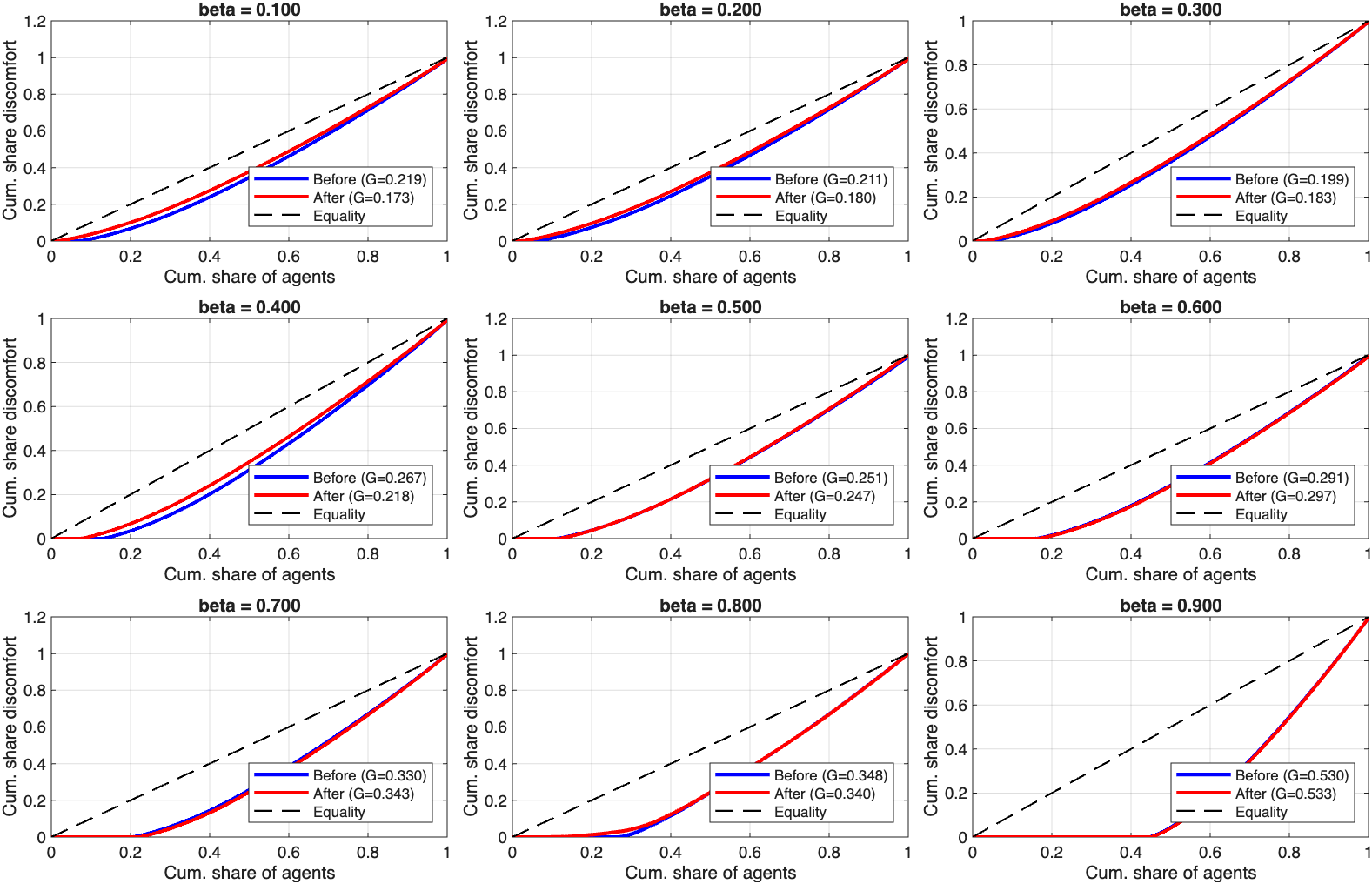}
		\caption{Lorenz curves illustrating the distribution of discomfort among agents for different values of the behavior parameter $\beta$. Each subplot compares the discomfort distribution before (blue) and after (red) applying the proposed slot exchange mechanism, with the dashed diagonal representing perfect equality. The corresponding Gini coefficients (G) are reported in each panel, quantifying inequality in discomfort distribution. Lower Gini values indicate fairer outcomes. The results show that the slot exchange mechanism generally reduces discomfort inequality for lower and moderate $\beta$ values, while limited improvement is observed when agents behave highly selfishly (high $\beta$).}
		\label{fig:ginniapendix}
	\end{figure}
	
	\begin{table*}[htbp]
		\centering
		\caption{Summary of Literature on MAS-based Energy Management and consumer Comfort Optimization}
		\resizebox{\textwidth}{!}{
			\begin{tabular}{p{1.5cm} p{2.5cm} p{3cm} p{3cm} p{3cm} p{3cm}}
				\toprule
				\textbf{Ref} & \textbf{Focus Area} & \textbf{Proposed Method} & \textbf{Key Techniques} & \textbf{Main Contributions} & \textbf{Outcomes} \\
				\midrule
				\cite{MAS}   & Smart Grid        & Agent-based solution for energy loss minimization & Storage system optimization & Reduce power losses \& costs & Improved consumer bills \\
				\cite{2}     & Multi-agent coordination & Graph-based policy factorization & Team coordination in dynamic environments & Better multi-agent cooperation & Validated in simulations \\
				\cite{MADC}  & Microgrids        & DMA-EMS (Decentralized) & Fault-tolerant dispatch, inter-microgrid power sharing & Reliable under faults & Enhanced reliability \\
				\cite{MARL}  & Home Energy Management  & Multi-agent RL framework & Load forecasting, consumer preferences & Cost reduction, prevent grid overloading & 15\% electricity cost savings \\
				\cite{MALS}  & Microgrids demand-side management    & MAS + Antlion Optimizer & Demand-side management & Minimize costs \& peak demand & Optimized distribution across sectors \\
				\cite{rw1}   & Energy Demand     & Multi-agent deep RL & Value decomposition, Q-networks, prioritized experience replay & Optimize demand, cost, comfort & Fast convergence, stable system \\
				\cite{rw3}   & DR                & Game theory-based DR & Evolutionary algorithm, payoff function & Load shifting, profit optimization & Price bidding \& optimal planning \\
				\cite{rw4}   & Demand-side management Energy Sharing & Distributed MAS & Prosumers’ coordination, dynamic pricing & Cooling load \& thermal storage mgmt & Reduced CO$_2$, operational cost, higher profit \\
				\cite{rw5}   & Peak Shaving      & MAS coordination & Transfer learning, K-means pre-learning & Peak shaving \& valley filling & Resolved latency \& blockage \\
				\cite{rw6}   & Economic Dispatch & MAS + PSO variants & JADE, dynamic pricing, incentives & Load \& gen. balance, emission scheduling & MA-Chaotic PSO: fastest convergence \\
				\cite{rw7}   & Smart Grids       & Two-stage robust framework & Simplex, IGWO & Multi-building energy savings & Simplex $>$ IGWO for peak load \\
				\cite{rw8}   & Hybrid Microgrids & MAS-based DR algorithm & Intelligent optimization, equitable decisions & DR for profit & Robust under changing strategies \\
				\cite{com1}  & Smart Home        & GA vs PSO for comfort & Kalman filter, sensor refinement & Max comfort, min energy & GA: 31\% less energy, 10\% comfort gain \\
				\cite{com2}  & Visual Comfort    & Negotiation-based model & Soft computing, dynamic adjustment & Multi-occupant profiles & Efficient adaptable lighting \\
				\cite{com3}  & Thermal Comfort   & Multi-objective optimization & Climate, air speed, humidity control & Educational buildings & 9.1 kWh/m$^2$ annual savings \\
				\cite{com4}  & indoor air quality \& Comfort    & Modified Bat Algorithm & Exponential inertia weight & Temp, light, indoor air quality optimization & Outperforms Firefly, ACO \\
				\cite{com5}  & Smart Home AI     & CFD + BPNN + AMOPSO-GWO & Hybrid AI, real-time control & Rapid prediction, adjustments & 35\% energy savings, better indoor air quality \\
				\cite{com6}  & Hydrogen Homes    & ENGO algorithm & Green hydrogen storage & Cost-comfort trade-off & Peak demand, cost reduction \\
				\cite{rr1} & EV charging & Cooperative pricing strategy & DLMP, Shapley value & Joint optimization of power and traffic-aware pricing & Reduced grid and road congestion \\
				\cite{rr2} & EV charging & Two-stage hierarchical model & Virtual pricing, Stackelberg game & Resolves conflicts among independent operators & Lower load fluctuation, less curtailment \\
				\cite{rr3} & Behavioral DR & Data-driven behavioral analysis & Serious games, flexibility modeling & Quantifies impact of behavioral flexibility & Higher self-consumption and fairness \\
				\cite{rr4} & Flexibility services& Stochastic bilevel optimization & Dynamic pricing, CVA & Compares aggregator vs direct control & Cost reduction depends on prosumer type \\
				\cite{rr5} & BTM flexibility strategies & Survey and taxonomy & P2P trading, RL, forecasting & Identifies architectures and regulatory gaps & Guidance for scalable BTM integration \\
				\cite{rr6} & P2P energy sharing & Hierarchical NEMS framework & Internal pricing, demand side management & Combines P2P sharing with demand flexibility & Cost and CO\textsubscript{2} reduction \\
				\cite{rr7} & Demand side management & Two-level MAS framework & Multi-agent RL & Integrates trading with production planning & Improved industrial cost efficiency \\
				\cite{rr8} & Electricity rights transfer & P2P quota exchange mechanism & Water-filling allocation & Fair redistribution of tariff quota & Improved billing fairness \\
				\cite{rr9} & DR & Literature review & Robust RL, distributed control & Identifies trends and open challenges & Research directions for scalable DR \\
				\cite{rr10} & Cloud scheduling & Metaheuristic scheduler & Genetic algorithms, local search & Improves convergence in high dimensions & Energy savings, faster scheduling \\
				\cite{rr11} & Energy scheduling & Hybrid heuristic solver & Local search, constraint handling & Efficient solutions for CECSP & High-quality feasible schedules \\
				\cite{rr12} & DR& Three-layer game-theoretic model & DR program design, games & Aligns DR incentives with system goals & Enhanced flexibility utilization \\
				\cite{rr13} & DR& Hierarchical DR framework & SAA, Stackelberg game & Manages uncertainty across timescales & Reduced imbalance, higher revenue \\
				\cite{rr14} & Demand side management & Noncooperative game formulation & Nash equilibrium analysis & Balances cost and privacy & Secure distributed optimization \\
				\cite{rr16} & Microgrid planning & Noncooperative planning model & Imperfect information games & Models rational MG investment decisions & Improved planning robustness \\
				\cite{rr17} & DR & Cooperative scheduling model & Cost-sharing settlement & Reduces demand charges via cooperation & Lower peak-related costs \\
				\cite{rr19} & DR & Iterative pricing mechanism & Decentralized optimization & Achieves scalable DR without central control & Improved load balance \\
				\hline
				\bottomrule
			\end{tabular}
		}
		\label{tab:literature_comparison}
	\end{table*}

\end{document}